%% file: 0_main.tex
\PassOptionsToPackage{prologue,table}{xcolor}
\documentclass[sigconf]{acmart}
\AtBeginDocument{%
  }

\newcommand{\bheading}[1]{\vspace*{.5em}\noindent{\textbf{#1.}}}

\copyrightyear{2025}
\acmYear{2025}
\setcopyright{acmlicensed}
\acmConference[CHI '25]{CHI Conference on Human Factors in Computing Systems}{April 26-May 1, 2025}{Yokohama, Japan}
\acmBooktitle{CHI Conference on Human Factors in Computing Systems (CHI '25), April 26-May 1, 2025, Yokohama, Japan}
\acmDOI{10.1145/3706598.3713564}
\acmISBN{979-8-4007-1394-1/25/04}

\usepackage{multirow}
\usepackage{array}

\definecolor{PastelGreen}{rgb}{0.8, 1.0, 0.8}    
\definecolor{PastelBlue}{rgb}{0.8, 0.88, 1.0}    
\definecolor{PastelOrange}{rgb}{1.0, 0.9, 0.8}   
\definecolor{PastelYellow}{rgb}{1.0, 1.0, 0.8}   

\begin{document}

\title{AI Suggestions Homogenize Writing Toward Western Styles and Diminish Cultural Nuances}

\author{Dhruv Agarwal}
\email{da399@cornell.edu}
\orcid{0000-0002-1090-3583}
\affiliation{%
  \institution{Cornell University}
  \country{USA}
}

\author{Mor Naaman}
\email{mor.naaman@cornell.edu}
\orcid{0000-0002-6436-3877}
\affiliation{%
  \institution{Cornell Tech}
  \country{USA}
}

\author{Aditya Vashistha}
\email{adityav@cornell.edu}
\orcid{0000-0001-5693-3326}
\affiliation{%
  \institution{Cornell University}
  \country{USA}
}

\renewcommand{\shortauthors}{Agarwal et al.}

\input{0_abstract}



\begin{teaserfigure}
  \centering
  \includegraphics[width=0.8\textwidth]{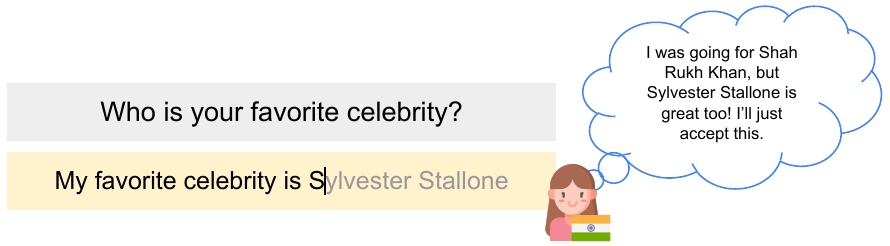}
  \caption{A representation of the potential cultural homogenization from Western-centric AI models}
  \Description{A screenshot of a writing task where the prompt asks, "Who is your favorite public figure?" The user begins typing, "My favorite public figure is S" and the autocomplete suggestion "Sylvester Stallone" is shown in grey. To the right, there is a cartoon of an Indian person, with a thought bubble saying, "I was going for Shah Rukh Khan, but Sylvester Stallone is great too! I’ll just accept this."}
  \label{fig:teaser}
\end{teaserfigure}


\maketitle

\section{Introduction} \label{sec:introduction}

\input{1_introduction}

\section{Related Work} \label{sec:related_work}

\input{2_related_work}

\section{Methodology} \label{sec:methodology}

\input{3_methodology}

\section{Findings} \label{sec:findings}

\input{4_findings}

\section{Discussion} \label{sec:fiscussion}

\input{5_discussion}

\section{Conclusion} \label{sec:conclusion}

\input{6_conclusion}

\begin{acks}
This work is supported by funding from Infosys, Microsoft AFMR Program, and Global Cornell. This material is based upon work partially supported by the National Science Foundation under Grant No. CHS 1901151/1901329. We also thank Anya Shukla for her insightful suggestions during the analysis.
\end{acks}

\bibliographystyle{ACM-Reference-Format}
\bibliography{references}

\appendix
\input{7_appendix}

\end{document}

%% file: 0_abstract.tex
\begin{abstract}
Large language models (LLMs) are being increasingly integrated into everyday products and services, such as coding tools and writing assistants. As these embedded AI applications are deployed globally, there is a growing concern that the AI models underlying these applications prioritize Western values. This paper investigates what happens when a Western-centric AI model provides writing suggestions to users from a different cultural background. We conducted a cross-cultural controlled experiment with 118 participants from India and the United States who completed culturally grounded writing tasks with and without AI suggestions. Our analysis reveals that AI provided greater efficiency gains for Americans compared to Indians. Moreover, AI suggestions led Indian participants to adopt Western writing styles, altering not just what is written but also how it is written. These findings show that Western-centric AI models homogenize writing toward Western norms, diminishing nuances that differentiate cultural expression.
\end{abstract}

%% file: 1_introduction.tex
Large language models (LLMs) are increasingly shaping online discourse and are being integrated into everyday applications to improve user productivity and experience. These include chatbots that aid in content comprehension\footnote{\href{https://www.adobe.com/acrobat/generative-ai-pdf.html}{AI Assistant in Adobe Acrobat}, \href{https://askyourpdf.com/}{AskYourPDF}}, code completion systems that streamline programming\footnote{\href{https://github.com/features/copilot}{GitHub Copilot}, \href{https://cursor.com/}{Cursor}}, and autocomplete tools that facilitate faster writing~\cite{Chen2019SmartCompose}. Writing, in particular, has become a prominent area where LLMs offer substantial assistance, supporting tasks such as scientific writing~\cite{gero2021sciencewrirting}, storytelling~\cite{Singh2023elephant}, and journalism~\cite{Petridis2023}. These applications provide inline writing suggestions, which users can accept or reject in real-time. Such AI writing suggestions have become integral to email clients (e.g., Gmail Smart Compose~\cite{Chen2019SmartCompose}), note-taking applications (e.g., Notion), and word processors (e.g., Google Docs). Their growing utility is evident from their integration into popular web browsers like Google Chrome and Microsoft Edge, which now offer autocomplete as a native feature, enabled by default\footnote{See announcements: \href{https://www.microsoft.com/en-us/edge/features/text-prediction}{Microsoft Edge}, \href{https://support.google.com/chrome/answer/14582048}{Google Chrome}}. 

While LLM-embedded applications like text autocomplete have gained widespread global adoption, these technologies also carry significant risks, such as perpetuating harmful stereotypes about marginalized groups and underserved communities. For instance, LLMs have been shown to depict Muslims as terrorists~\cite{abid2021persistentantimuslimbiaslarge} and reinforce negative stereotypes about disabled individuals~\cite{venkit-etal-2022-study, gadiraju_i_2023}.
Additionally, research has revealed that LLMs prioritize Western norms and values in their interactions~\cite{johnson2022ghost, cao2023chatgptculture}, resulting in representational harms for diverse non-Western cultures~\cite{Shelby2023, basu2023inspecting, Qadri2023regimes}.

While explicit cultural stereotyping shown by prior work is deeply problematic, an even more insidious issue lies in the subtle biases LLMs can introduce through their suggestions. For example, recent work shows that autocomplete suggestions can change users' language~\cite{Hohenstein2023language}, writing~\cite{Poddar2023topics}, and even attitudes about social topics~\cite{Jakesch2023, WilliamsCeci2024}. Furthermore, unlike chat-based applications such as ChatGPT, LLM-embedded applications do not allow users to prompt and fine-tune the model to suit their cultural preferences. This creates situations where users and AI may have different cultural values, potentially leading to conflicts over whose norms should be expressed.
While prior studies have explored cultural harms in LLMs using open-ended prompts, no research has yet examined such cultural clashes in embedded AI applications. In this paper, we examine how AI suggestions influence user-generated content when the AI and users share the same cultural identity versus when they do not. In particular, we ask: 

\begin{itemize}
  \item[\textbf{RQ1}:] Does writing with a Western-centric AI provide greater benefits to users from Western cultures, compared to those from non-Western cultures?
  \item[\textbf{RQ2}:] Does writing with a Western-centric AI homogenize the writing styles of non-Western users toward Western styles?
\end{itemize}

To answer these questions, we conducted a cross-cultural experiment with $118$ users from India and the US recruited through Prolific, an online crowdsourcing platform. Participants from both cultures were asked to complete writing tasks in English. We designed the task prompts using Hofstede's Cultural Onion framework~\cite{Hofstede:1991}, which allowed us to elicit cultural practices ranging from explicit (e.g., food) to implicit (e.g., rituals). Each participant was randomly assigned either to the AI condition receiving autocomplete suggestions from GPT-4o while writing or the No AI condition writing organically without AI assistance.
Subsequently, we compared the essays written by participants in the four experimental groups (Indian and American users writing with and without AI suggestions).

Our analysis yielded two key results. In response to RQ1, we found that while AI boosts productivity for both Indian and American participants, the gains are higher for American participants. This discrepancy raises concerns about quality-of-service harms~\cite{Shelby2023} for non-Western users, wherein they need to put in more effort to achieve similar benefits. In response to RQ2, we found that
AI caused Indian participants to write more like Americans, thereby homogenizing writing toward Western styles and diminishing nuances that differentiate cultural expression. Worryingly, AI influences not just \textit{what} is written (e.g., shifting preferences toward Western cultural artifacts such as food items), but also more ingrained elements of \textit{how} it's written, silently erasing non-Western styles of cultural expression. For instance, with AI suggestions, Indians lose cultural nuance and describe their own food and festivals from a Western gaze. We discuss the implications of these cultural harms from a colonial lens and propose strategies to mitigate them.
Overall, we make the following contributions to the HCI and AI communities:
\begin{itemize}
    \item We present the first cross-cultural analysis of AI writing suggestions, providing concrete evidence of higher productivity gains for Western users compared to non-Western users. 
    \item We show how cultural homogenization occurs when non-Western users write with Western-centric AI models. 
    \item We discuss the potential harms of cross-cultural homogenization and propose strategies to address the harms of cultural imperialism and linguistic singularity.
\end{itemize}

%% file: 2_related_work.tex
We first review scholarly work exploring cultural bias in AI models, emphasizing how LLMs prioritize Western norms and values. Given our focus on AI-based writing suggestions, we then situate our work within the emerging HCI scholarship on designing, developing, and evaluating AI technologies for writing support. 

\subsection{Cultural Bias in AI Models}
An emerging body of HCI and AI scholarship has examined the cultural composition of generative AI technologies~\cite{adilazuarda2024measuring}, and found that these models center Western norms and values~\cite{cao2023chatgptculture, Tao2024}. For example, \citet{Qadri2023regimes} showed that text-to-image models fail to generate cultural artifacts, amplify hegemonic cultural defaults, and perpetuate cultural tropes when depicting non-Western cultures. Even \emph{within} the West, these AI models tend to prioritize US-centric values and artifacts~\cite{basu2023inspecting, Schwbel2023, venkit-etal-2022-study, johnson2022ghost}. Furthermore, LLMs trained and probed in non-Western languages (e.g., Arabic) continue to show Western bias~\cite{naous2024beerprayer, arora2023hofstedellm}. Recognizing these biases, recent work has proposed benchmark datasets to evaluate the multicultural knowledge of LLMs~\cite{chiu2024culturalbench, rao2024normad}.
Together, these studies reveal the harms of culturally incongruent AI models~\cite{prabhakaran2022cultural}. They underscore how data sourced from low-paid workers in non-Western regions is commodified to create models that inadequately serve these regions, a phenomenon known as data colonization~\cite{couldry2019datacolonialism}. Moreover, these systems reinforce Western cultural hegemony, perpetuate cultural erasure, and amplify stereotypes~\cite{Qadri2023regimes}, echoing the dynamics of colonial times and giving rise to AI colonialism.~\cite{Hao2022aicolonialism, tacheva_ai_2023}.


Addressing these cultural biases is crucial not only to prevent representational, allocative, and quality-of-service harms~\cite{barocas2017problem, Shelby2023, Anuyah2023} but also to avoid culture clashes that arise when the values embedded in AI models diverge from those of their users~\cite{prabhakaran2022cultural, hershcovich-etal-2022-challenges}. Early work has acknowledged these challenges of value plurality, questioning what it means to imbue AI with human values when those values vary across cultures~\cite{tamkin2021understanding}, and which values a model should prioritize when producing a singular output~\cite{johnson2022ghost}. 
As generative AI technologies like LLMs are increasingly integrated into global products and services, these questions have become critically important.

The commercial growth has also driven a shift toward embedded AI applications, where the underlying models are obscured from users. This opacity makes it challenging to mitigate cultural biases through open-ended prompting or fine-tuning, thereby surfacing cultural clashes. We examine this culture clash that arises when users interact with embedded AI models that are culturally distant from them. To our knowledge, this is the first to reveal cultural homogenization effects when embedded AI features continue to be powered by culturally biased models.

\subsection{AI-Based Writing Support}

Advances in computing have been extensively used to provide writing support, starting with tools like spell-check~\cite{Peterson1980spellcheck} and grammar-check~\cite{Leacock2010grammar} to improve writing efficiency~\cite{Buschek2021}. As natural language generation technologies evolved, researchers began applying them to support creative writing pursuits such as story writing~\cite{Dhillon2024}. Some examples include providing next-word or next-sentence suggestions~\cite{Clark2018, Manjavacas2017, Roemmele2018}, and generating creative content (e.g., poems) based on a given topic or writing style~\cite{Ghazvininejad2016poetry, Gabriel2015}. However, early AI models struggled to capture the intent of the writer, rendering these tools as passive ideation tools rather than active writing assistants~\cite{Gero2019metaphoria} (e.g., metaphor generation~\cite{Gero2019thesaurus, Chakrabarty2021metaphor}).

However, recent advancements in natural language generation, driven by LLMs, have led to a resurgence of tools to support open-ended writing~\cite{Dhillon2024}, such as 
story generation~\cite{Singh2023elephant, Lee2022, Yang2022AIAA, Yuan2022wordcraft} and screenplay writing~\cite{Mirowski2023}. Beyond creative writing, HCI researchers have also proposed tools for argumentative writing, including journalism~\cite{Petridis2023}, scientific writing~\cite{gero2021sciencewrirting}, email writing~\cite{Kannan2016}, and crafting self-introductions for networking~\cite{Hui2018help}. These tools are designed to work in situ, as active collaborators, offering writing suggestions through pop-ups, side panels, or inline predictions~\cite{Chen2019SmartCompose}. 

With the popularity of such tools, researchers have also examined the broader implications of AI-powered writing suggestions, including their effect on users and the resulting text~\cite{Buschek2021}. For instance, \citet{Arnold2016} identified a trade-off between efficiency and ideation: while short single-word suggestions enhance efficiency without inspiring creativity, long multi-word suggestions offer inspiration but can be distracting. As a result, studies have shown mixed results regarding the efficiency benefits of writing suggestions~\cite{Dhillon2024, Quinn2016, Bhat2023Interacting}, indicating that such gains may be context-dependent. Despite the potential to improve quality and productivity, suggestions often lead to reduced user satisfaction and a loss of ownership over the text produced~\cite{Dhillon2024, kadoma2024inclusion}.

Furthermore, there has been growing interest in the risks of writing support tools, especially as they are powered by language models that reflect social biases. For instance, \citet{Jakesch2023} showed that co-writing with opinionated language models shifts users' views to align with the model's biases. \citet{kadoma2024inclusion} highlighted the differential impact of writing assistants on users from minoritized genders. We extend this line of work by presenting the first cross-cultural study of writing suggestions, specifically designed to measure the cultural harms that may arise from these tools. The most closely related work in this space is \citet{Buschek2021}, who investigated the differential impact of multi-word suggestions on native and non-native English speakers. However, their study focused on a coarse-grained comparison of English language proficiency without controlling for cultural differences. In contrast, we present a systematic study that explicitly investigates cross-cultural differences in the context of writing suggestions from culturally biased models.

%% file: 3_methodology.tex
We conducted a controlled experiment with $118$ participants, comprising $60$ Indian and $58$ American users, recruited through the online crowdsourcing platform Prolific. Participants completed short writing tasks designed to elicit cultural values and artifacts. We had a standard $2\times2$ study design in which participants from India and the US completed the tasks with or without AI suggestions. Subsequently, we compared the essays written by participants from the four groups. Below, we describe our methodology in detail.

\subsection{Experiment Design}

\input{tables/experimental_groups}

To examine the impact of writing with a culturally incongruent model, we simulated a ``cultural distance'' between the users and the model. Since AI models are often aligned with Western cultures and values~\cite{johnson2022ghost, Qadri2023regimes, Moayeri2024WorldBench, basu2023inspecting}, we conducted the study with participants from the US (representing a smaller cultural distance) and India (representing a larger cultural distance). Participants from both cultures were randomly assigned to complete writing tasks with or without AI suggestions, resulting in a standard $2\times2$ between-subjects study design. Hence, each participant was assigned to one of four experimental groups: Indians writing with or without AI, and Americans writing with or without AI\footnote{For clarity, we use the term ``group'' to refer to one of the four experimental groups, ``condition'' to refer to the AI condition (AI vs No AI), and ``cohort''/``culture''/``country'' to refer to the cultural group (Indian vs American participants).}. These groups are summarized in Table~\ref{tab:experimental_groups}.

The two control groups (without AI) helped us capture the natural writing styles of participants from both cultures, forming a baseline for comparison with the treatment groups. The treatment groups (with AI) were designed to investigate whether writing styles changed when AI suggestions were introduced.

\subsection{Writing Tasks}

Each participant was required to complete four writing tasks in English. Although English is not a native language for most Indians, we expected our participants to be fluent due to being on Prolific, a platform run entirely in English. Indeed, 59 of our 60 Indian participants reported speaking English. The task topics were designed to elicit various aspects of culture defined by Hofstede in his ``Cultural Onion''~\cite{Hofstede:1991} (Figure~\ref{fig:cultural_onion}). Hofstede's conceptualization of culture, such as the six cultural dimensions~\cite{hofstede2001culture}, and the cultural onion~\cite{Hofstede:1991}, has been widely used in prior work to operationalize culture~\cite{Lu2022, Guo2022, arora2023hofstedellm}. In particular, the cultural onion framework uses the metaphor of a layered onion to model culture, envisioning an outsider progressing through layers of explicit cultural practices (\textit{symbols}, \textit{heroes}, \textit{rituals}) to ultimately grasp the core of the culture, its implicit \textit{values}. This layered approach allows us to elicit both explicit cultural practices and implicit values.

Specifically, \textit{symbols} are words and objects that hold a certain meaning in a culture, including language, art, clothes, food, etc. Symbols are not fixed and evolve but are observable by outsiders to the culture. \textit{Heroes} are people idolized by a culture, often because they possess characteristics that are valued within the culture. They may be real (e.g., APJ Abdul Kalam in India) or fictional (e.g., Rocky Balboa in the US). \textit{Rituals} are collective activities deemed important by the people of the culture. This may include holidays and festivals, ways of greeting, or other social customs (e.g., sauna). These three layers are collectively referred to as \textit{practices} and can be observed and practiced by outsiders even without necessarily understanding their underlying cultural meanings. Finally, at the core of the onion are the \textit{values}, which are preferences for certain states of affairs (e.g., individualism or collectivism). They are static and so ingrained that even people within the culture may be unaware of them. 

To holistically evoke these cultural nuances, we designed four writing tasks for our study, one for each layer of the cultural onion. The task prompts are shown in Table~\ref{tab:task_prompts}. To select these prompts, we chose representative artifacts from the outer layers of the onion and designed a relatable digital interaction (writing an email to a superior) to evoke underlying cultural values. There was an additional task in the study to function as an attention check. Attention checks are commonly used tools to improve the quality of data obtained from crowdsourcing platforms~\cite{Hauser2015, Kung2017}.
We introduced an attention check midway through the study (after two writing tasks), where the prompt instructed participants to leave the textbox blank. None of our participants failed the attention check.

\input{tables/essay_topics_and_onion}

\subsection{Study Procedure}
We published the study on Prolific, a widely used crowdsourcing platform. The study consisted of two parts: completing the writing tasks on our study portal, and subsequently filling out a demographic survey. 
Participants had to complete both parts to complete the study and receive payment. Below, we describe these two parts of the study.

\subsubsection{Study Portal}

\begin{figure*}[t]
    \centering
    \begin{tabular}{c}
    \includegraphics[height=2.5in]{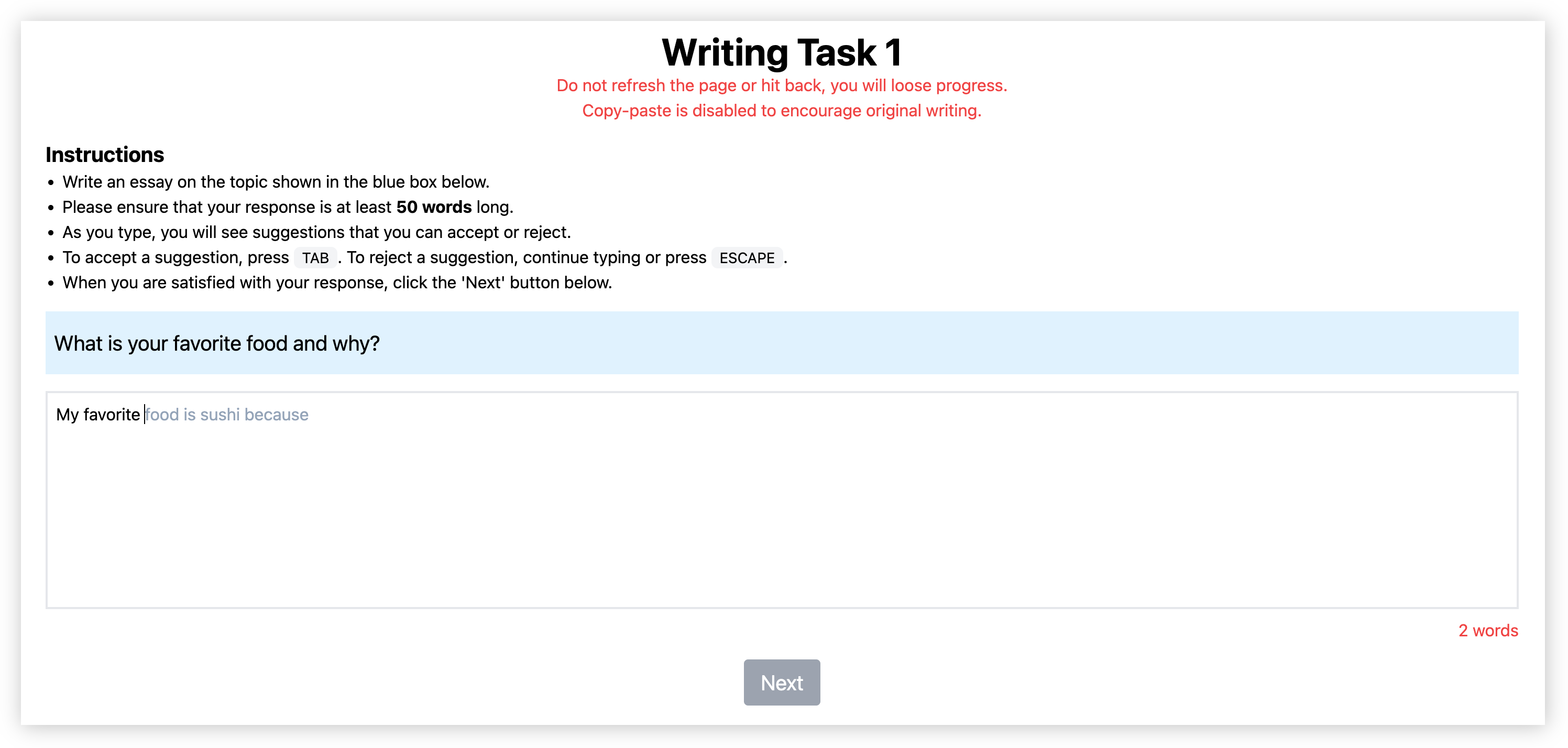}
    \end{tabular}
    \caption{The interface to complete the writing tasks in the AI condition. Suggestions were shown in grey next to the text written by the user and could be accepted by pressing the TAB key.}
    \Description{Screenshot of a writing task interface with a text box where users type a response to the prompt "What is your favorite food and why?" The user has begun typing "My favorite " and an inline suggestion, "food is sushi because" is shown in grey. Instructions are displayed at the top, including word count requirements and key bindings for accepting or rejecting suggestions. There is a word counter at the bottom right, and a greyed-out Next button at the bottom center.}
    \label{fig:writing_screenshot}
\end{figure*}

The task instructions on Prolific directed participants to an external study portal, which we built using React and hosted at a public link. 
The landing page of the portal welcomed participants and provided a brief description of the study. It also displayed a link to the informed consent form and a check box for providing consent. Participants who provided consent were randomly assigned to the AI or No AI condition. Those in the AI condition were then directed to a tutorial, which was a pop-up-style walkthrough demonstrating how to accept or reject AI suggestions. After participants accepted and rejected a few suggestions, they proceeded to the main writing tasks.

Each participant needed to complete four writing tasks (plus the attention check), presented sequentially. The interface for each task was the same and is shown in Figure~\ref{fig:writing_screenshot}. At the top, there were instructions. We enforced a minimum word requirement (50 words) to encourage participants to engage meaningfully with the task prompt~\cite{Buschek2021, Jakesch2023}. For the AI condition, the instructions included guidance on accepting or rejecting suggestions; we provided these instructions as a reminder despite the tutorial at the beginning.

Below the instructions was a textbox for their response. In the AI condition, the textbox fetched an inline autocomplete suggestion from GPT-4o if the user paused typing for 100 ms. While some related work used longer delays~\cite{Jakesch2023, Buschek2021}, participants in those studies complained of long loading times~\cite{Buschek2021}. Hence, we chose a shorter delay, especially as the model itself took $\sim$500 ms to respond with a suggestion. To prevent participants from rapidly accepting suggestions to speed through the study, we required participants to type at least one character after accepting a suggestion before another suggestion was shown. This approach was intended to encourage participants to engage more thoughtfully with the task, rather than relying on AI suggestions to guide their next chain of thought. Suggestions were displayed in light grey as autocomplete suggestions next to the text written by the user (similar to smart completion on Gmail or VS Code). The user could press the TAB key to accept a suggestion, the escape key to explicitly reject a suggestion, or continue typing to implicitly ignore a suggestion. We logged all these interactions in an external database.

Below the textbox, a word counter offered real-time feedback on how many more words were needed to meet the minimum requirement. Once the minimum requirement was met, the word counter turned green and the ``Next'' button was activated.

\subsubsection{Retrieving Autocomplete Suggestions} \label{subsubsec:retrieving_autocomplete_suggestions}
We used the following prompt to retrieve autocomplete suggestions from OpenAI's latest language model at the time, GPT-4o. We refined this prompt through iterative prompt engineering until the research team was satisfied with the generated suggestions.
\begin{quote}
\texttt{You are an AI autocomplete assistant. You need to provide short autocomplete suggestions to help people with writing. Some guidelines:\\
- Your suggestion should make sense inline (it will be shown to the user as ghost text).\\
- If the user has just completed a word, add a space before the suggestion.\\
- Suggestions should be <=10 words\\
- The user is writing about the topic: "<task prompt from Table~\ref{tab:task_prompts}>"\\
- Output a JSON of the following format: \{"suggestion": "<your suggestion here>"\}}
\end{quote}
We embedded the essay topic in the LLM prompt to provide contextually relevant suggestions to users~\cite{Jakesch2023, Buschek2021}. We also considered explicitly directing the model to generate Western-oriented suggestions. While this approach would enhance the robustness of our study by controlling the model's behavior, it would make it less realistic as users may not specify such culturally biased prompts in the real world. To explore this, we performed a preliminary evaluation. We added the instruction ``Give American suggestions'' to the above prompt and observed the model's output with and without this additional instruction. We found that with both prompts, the model (a) defaulted to Western suggestions, and (b) adapted to Indian suggestions when the leading text provided Indian context. This behavior aligns with research showing that AI models default to Western norms and values~\cite{johnson2022ghost, Qadri2023regimes}. Hence, we allowed the model to operate neutrally to improve the real-world applicability of our findings.

\subsubsection{Demographic Survey}
After completing all the writing tasks, participants were directed to a demographic survey via a unique link. To avoid biasing their responses, the survey was intentionally placed at the end of the study. This was a two-page survey designed on Qualtrics asking participants their age, gender, country of birth and residence, years lived in the country, education level, occupation, and languages spoken. These demographic details were used both to triangulate participants' cultural backgrounds (see Section~\ref{subsec:participants}) and to ensure diversity \textit{within} each culture on axes such as age and gender.

We selected India and the US for this study due to their different cultural compositions. However, since Prolific predominantly operates in the West, our Indian participants might be more familiar with American culture than the average Indian, due to their exposure to international crowdsourcing tasks. This could reduce the cultural distance between the AI model and our participants, potentially skewing the generalizability of our findings. To ensure cultural differences between participants from India and the US, we included the Short Schwartz Value Survey (SSVS)\cite{Lindeman2005} in the demographic survey. The SSVS measures individual and cultural differences across ten values, such as power, achievement, tradition, and hedonism, and is a condensed version of the widely used Schwartz Value Survey\cite{Schwartz1992}.

Upon completing the survey, the participants were redirected back to Prolific to mark the study as complete. The research team verified that participants completed both parts of the study (the writing tasks and the demographic survey) before approving the payment (\$2 per participant).

\subsection{Participants} \label{subsec:participants}

\input{tables/participant_demographics}
We used Prolific to recruit identity-verified human crowdworkers. We included participants over 18 years old (a requirement for all Prolific users) residing in either the US or India. However, the current country of residence alone was not sufficient to map a participant to a specific culture. Hence, we retained only those participants who were born in and had lived their entire lives in the same country. This was confirmed using data from the demographic survey including age, country of birth and residence, and years lived in the country.

In total, we collected responses from $118$ participants: $60$ from India and $58$ from the US. The scale of our study was limited by the small number of Indian participants on Prolific. To get equal participation across the two cultures, we initially launched the study in India and allowed the participant count to stabilize before extending the study to the same number of participants in the US. The demographic details of the participants are summarized in Table~\ref{tab:participant_demographics}.

As discussed previously, we also administered the Short Schwartz Value Survey (SSVS) to our participants to measure their cultural differences. Indian and American participants had significantly different scores ($p < 0.05$ on an independent samples t-test) on eight of the ten values measured by the survey (e.g., power, hedonism, tradition, conformity, etc.). The two values with no significant difference were benevolence and self-direction. 
The score difference on eight of the ten values suggests significant cultural differences between participants in India and the US. 

\subsection{Analysis} \label{subsec:analysis}
We quantitatively analyzed the data collected from our online experiment by comparing task-level metrics across the four experimental groups: Indians and Americans, both with and without AI. We analyzed two sets of data: interaction logs captured on the study portal, and the final essays written by the participants. Below, we summarize the metrics used in our analyses.

\subsubsection{Interaction Logs}
The portal recorded data such as the time taken to complete a task and the number of suggestions seen, accepted, and rejected. Based on this data, we defined the following metrics.

\begin{enumerate}
    \item \textbf{Suggestion Acceptance Rate:} This measures the proportion of suggestions accepted in a task~\cite{Buschek2021}. Thus, it quantifies engagement with the AI suggestions. It is computed as follows:
    \begin{equation*}
        \text{Acceptance Rate} = \frac{\text{Number of suggestions accepted}}{\text{Number of suggestions shown}}
    \end{equation*}
    It ranges from $0$ to $1$, where $0$ means that no suggestion was accepted in completing a task. Thus, it can be interpreted as the percentage of suggestions accepted in a task. However, this is a coarse measure as a user may modify or delete a suggestion after accepting it. The next two measures account for this possibility.

    \item \textbf{AI Reliance:} This metric quantifies the proportion of the final essay generated by AI~\cite{kadoma2024inclusion, Buschek2021}. For each character in the final essay, we determine whether it originated from an AI suggestion or was authored directly by a participant. To handle complex cases when an accepted suggestion is modified, we use the longest common subsequence (LCS) method so that only the retained portion of the suggestion contributes to AI reliance.\footnote{While this method may underestimate AI reliance in cases where modifications occur midway through a suggestion, it still provides a lower-bound approximation for analysis. This conservative approach ensures the reliability and robustness of our results.} AI reliance, thus, for each task is calculated as follows:
    \begin{equation*}
        \text{AI reliance} = \frac{\text{Number of AI-suggested chars}}{\text{Total number of chars in the essay}}
    \end{equation*}
    This metric ranges from $0$ to $1$. Lower values indicate lesser reliance on AI (i.e., greater human contribution), while higher values indicate a stronger reliance on AI-generated content. Unlike the acceptance rate, this metric is sensitive to suggestions that were accepted but later modified or deleted.

    \item \textbf{Suggestion Modification:} Users may accept a suggestion and modify it to better suit their requirements. For each task, we computed a boolean indicating whether a suggestion was modified during that task. We measure the \textit{existence} of a modification, rather than the rate or frequency~\cite{Buschek2021} because as little as one suggestion modification early in the essay can make subsequent suggestions culturally relevant. To capture modification, we checked if an accepted suggestion appeared in the final essay. If it was not found as-is, we assumed it was modified (or deleted). 

    \item \textbf{Writing Productivity:} Productivity is multi-dimensional (e.g., quality, readability, cognitive load), though in this work we define it as the number of words written by the user per second~\cite{Dhillon2024}. It can also be interpreted as the typing speed of the participant. 
    \begin{equation*}
        \text{Productivity} = \frac{\text{Number of words in the essay}}{\text{Time taken to finish the essay (in seconds)}}
    \end{equation*}
\end{enumerate}

\subsubsection{NLP Metrics} We also used established techniques from the NLP literature to analyze the essays written by the participants.

\begin{enumerate}
    \item \textbf{Type-Token Ratio (TTR):} TTR is a common measure of lexical diversity in text~\cite{Baker2006}. It is calculated as:
    \begin{equation*}
        \text{TTR} = \frac{\text{Number of unique words (types)}}{\text{Total number of words (tokens)}}
    \end{equation*}
    The value is bounded between 0 and 1; a low TTR represents more repetition of words and therefore less linguistic variation, and a high TTR shows more linguistic variation. Note: Although TTR penalizes longer texts, this was not an issue in our analysis, as the essays across all four experimental groups were of similar length, likely due to the minimum word requirement we enforced.
    
    \item \textbf{Cosine Similarity:} We fetched 3072-dimensional embeddings for each essay from OpenAI's text embedding model. We used these embeddings to calculate similarity scores between pairs of essays by computing the cosine similarity. Cosine similarity ranges between -1 and 1, with more negative values representing dissimilar texts and positive values representing similar texts.
\end{enumerate}

\noindent In addition to these quantitative metrics, we also conducted a content analysis of the essays. This involved manually coding the essays to identify references to cultural symbols and idols, as well as examining instances of exoticization, misrepresentation, and Western gaze~\cite{Qadri2023regimes}. This qualitative lens provided additional insights into the cultural shifts observed in the AI and No AI conditions.

\subsection{Ethics and Positionality}
This study was approved by the IRB of our institution, which included a careful review of the risks of exposing study participants to generative AI applications. Since culture is closely tied to one's identity, we were careful in designing a culturally appropriate study. For example, the third essay prompt asked the participants to write about their favorite holidays, but the equivalent term in India was ``festivals'', so we framed the prompt to include both terms. 

Our team has multiple members who have lived in both India and the US. This exposure to both cultures provided us with valuable context for designing this study and interpreting its results. Finally, we approached this work from an action research mindset, aiming to study the harms of working with culturally incongruent AI models. Building on insights generated from this work, in the future we aim to take action toward reducing these harms.


%% file: tables/experimental_groups.tex
\begin{table}[t]
\begin{tabular}{|l|l|l|}
\hline
\multicolumn{1}{|c|}{\textbf{Type}} & \multicolumn{1}{c|}{\textbf{Description}} & \multicolumn{1}{c|}{\textbf{Participants}} \\ \hline
\multirow{2}{*}{Control} & Indians writing without AI & 24 \\ \cline{2-3} 
 & Americans writing without AI & 29 \\ \hline
\multirow{2}{*}{Treatment} & Indians writing with AI & 36 \\ \cline{2-3} 
 & Americans writing with AI & 29 \\ \hline
\end{tabular}
\caption{Four experimental groups in our study.}
\Description{A table summarizing the four experimental groups in the study. The "Control" group includes 24 Indians and 29 Americans writing without AI. The "Treatment" group includes 36 Indians and 29 Americans writing with AI suggestions.}
\label{tab:experimental_groups}
\end{table}

%% file: tables/essay_topics_and_onion.tex

\begin{figure}[t]
    \centering
    \includegraphics[height=1.71in]{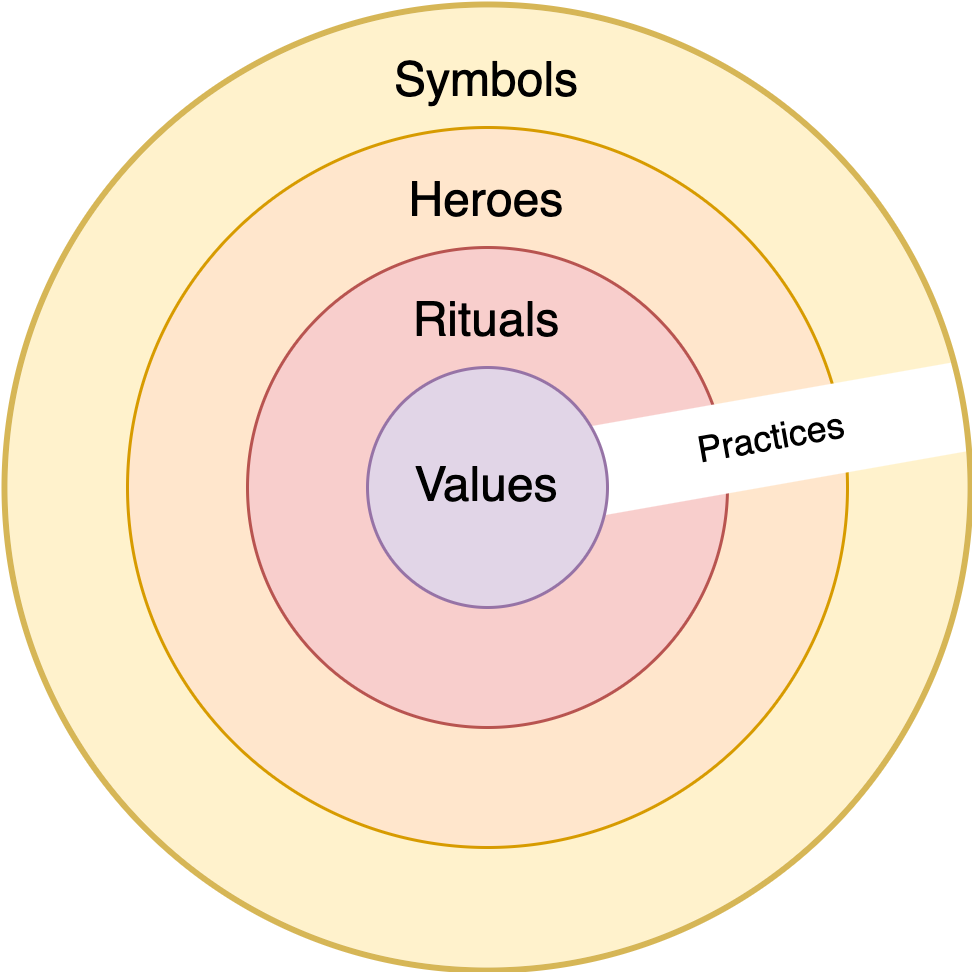}
    \caption{Hofstede's Cultural Onion}
    \Description{A series of four concentric circles representing layers of culture. The innermost circle is labeled "Values," followed by "Rituals," "Heroes," and "Symbols." A diagonal label reading "Practices" crosses all outer layers (except the Values core), indicating that the outer elements manifest in cultural practices.}
    \label{fig:cultural_onion}
\end{figure}

\begin{table}[t]
\centering
\begin{tabular}{|l|l|}
\hline
\multicolumn{1}{|c|}{\textbf{Layer}} & \multicolumn{1}{c|}{\textbf{Task Prompt}} \\ \hline
Symbols & \begin{tabular}[c]{@{}l@{}}What is your favorite food\\ and why?\end{tabular} \\ \hline
Heroes & \begin{tabular}[c]{@{}l@{}}Who is your favorite celebrity\\ or public figure and why?\end{tabular} \\ \hline
Rituals & \begin{tabular}[c]{@{}l@{}}Which is your favorite\\ festival/holiday and how do\\ you celebrate it?\end{tabular} \\ \hline
Values & \begin{tabular}[c]{@{}l@{}}Write an email to your boss\\ asking them for a two-week\\ leave with information about\\ why you need to be away.\end{tabular} \\ \hline
\end{tabular}
\captionof{table}{Prompts given to participants for the four writing tasks. These topics were designed to elicit various aspects of culture as defined by Hofstede's cultural onion.}
\Description{A table outlining the writing prompts provided to participants, based on Hofstede’s cultural onion layers. For the "Symbols" layer, participants were asked, "What is your favorite food and why?" For "Heroes," they were asked, "Who is your favorite public figure and why?" For "Rituals," the prompt was, "Which is your favorite festival/holiday and how do you celebrate it?" Lastly, for "Values," participants were asked to "Write an email to your boss asking for a two-week leave with information about why you need to be away."}
\label{tab:task_prompts}
\end{table}

%% file: tables/participant_demographics.tex
\begin{table*}[t]
\centering
\begin{tabular}{|l|l|l|l|l|l|}
\hline
\multicolumn{1}{|c|}{\textbf{}} &
  \multicolumn{1}{c|}{\textbf{Age (years)}} &
  \multicolumn{1}{c|}{\textbf{Gender}} &
  \multicolumn{1}{c|}{\textbf{Education}} &
  \multicolumn{1}{c|}{\textbf{Languages}} &
  \multicolumn{1}{c|}{\textbf{Occupations}} \\ \hline
\textbf{\begin{tabular}[c]{@{}l@{}}India\\ ($n$=60)\end{tabular}} &
  33.38 $\pm$ 11.85 &
  \begin{tabular}[c]{@{}l@{}}Male: 71.7\%\\ Female: 26.7\%\\ Undisclosed: 1.7\%\end{tabular} &
  \begin{tabular}[c]{@{}l@{}}Masters: 41.7\%\\ Bachelors: 43.3\%\\ High-School: 15\%\end{tabular} &
  18 unique &
  18 unique \\ \hline
\textbf{\begin{tabular}[c]{@{}l@{}}US\\ ($n$=58)\end{tabular}} &
  36.07 $\pm$ 13.52 &
  \begin{tabular}[c]{@{}l@{}}Male: 48.3\%\\ Female: 50\%\\ Non-binary: 1.7\%\end{tabular} &
  \begin{tabular}[c]{@{}l@{}}Masters: 50\%\\ Bachelors: 39.7\%\\ High-School: 8.6\%\\ No school: 1.7\%\end{tabular} &
  10 unique &
  30 unique \\ \hline
\end{tabular}
\caption{Self-reported participant demographics. The full list of languages and occupations is given in Appendix~\ref{appendix:participant_demographics}.}
\label{tab:participant_demographics}
\Description{A table displaying self-reported participant demographics for Indian (n=60) and US (n=58) participants. Age (years): Indian participants have an average age of 33.38 ± 11.85, while US participants have an average age of 36.07 ± 13.52. Gender: For India, 71.7\% are male, 26.7\% female, and 1.7\% undisclosed. For the US, 48.3\% are male, 50\% female, and 1.7\% non-binary. Education: In India, 41.7\% hold a master's degree, 43.3\% a bachelor's, and 15\% completed high school. In the US, 50\% hold a master's degree, 39.7\% a bachelor's, 8.6\% completed high school, and 1.7\% reported no schooling. Languages: Indian participants speak 18 unique languages, while US participants speak 10 unique languages. Occupations: Indian participants report 18 unique occupations, while US participants report 30 unique occupations.}
\end{table*}

%% file: 4_findings.tex
Overall, $118$ participants completed four tasks each, resulting in a dataset of $472$ essays. Among these participants, $65$ received AI-generated suggestions to aid their writing---$36$ in India and $29$ in the US. These participants saw a total of 12,015 suggestions, accepting 1,476 of them ($12.3\%$). Of the rejected suggestions, only $0.6\%$ were explicitly rejected by pressing the escape key; the rest were dismissed by continuing to type on. Further, a majority of the rejected suggestions ($68.8\%$) were dismissed within 500 ms of appearing on the screen, indicating that they were dismissed in the flow of writing, perhaps without being fully considered by the user. These phases represent bursts of focused manual typing where AI suggestions may not be desired~\cite{Buschek2021}. The most common suggestions provided by the model are shown in Table~\ref{tab:top_suggestions}; there is a noticeable Western bias in the suggestions for the food and festival tasks.

\input{tables/top_trigrams}

Participants were more likely to accept suggestions as the task progressed: of the suggestions they saw in the first third of the task, they accepted $9.2\%$, in the second third they accepted $12.1\%$, and in the final third of the task they accepted $15.9\%$. This may be because suggestions become more useful with more context. In the rest of this section, we present a cross-cultural analysis of the data.

\input{4.1_}

\input{4.2_}

\input{4.3_}

\input{4.4_}

%% file: tables/top_trigrams.tex
\begin{table*}[t]
\centering
\begin{tabular}{|l|l|}
\hline
\multicolumn{1}{|c|}{\textbf{Task Prompt}} & \multicolumn{1}{c|}{\textbf{Top Trigrams}} \\ \hline
Food* & is pizza because; food is pizza; favorite food is; is sushi because \\ \hline
Public figure & because of his; favorite celebrity is; known for his \\ \hline
Festival/holiday* & is christmas because; festival is christmas; family and friends \\ \hline
Email for vacation & a two-week leave; finds you well; request a two-week \\ \hline
\end{tabular}
\caption{Top autocomplete suggestions (trigrams) provided by the model for each task prompt. Suggestions for the food and festival tasks (marked with an asterisk) show a noticeable Western bias, while suggestions for the other two tasks are more generic.}
\label{tab:top_suggestions}
\Description{A table displaying the top trigram autocomplete suggestions provided by the model for each task prompt. Food*: Suggestions include "is pizza because," "food is pizza," "favorite food is," and "is sushi because." Public figure: Suggestions include "because of his," "favorite celebrity is," and "known for his." Festival/holiday*: Suggestions include "is Christmas because," "festival is Christmas," and "family and friends." Email for vacation: Suggestions include "a two-week leave," "finds you well," and "request a two-week." Suggestions for the food and festival tasks (marked with an asterisk) show a noticeable Western bias, while the suggestions for the other two tasks are more generic.}
\end{table*}

%% file: 4.1_.tex
\subsection{Does AI impact Indians and Americans differently?} \label{subsec:differential_impact}

\begin{figure*}[t]
    \centering
    \begin{tabular}{ccc}
    \includegraphics[height=1.47in]{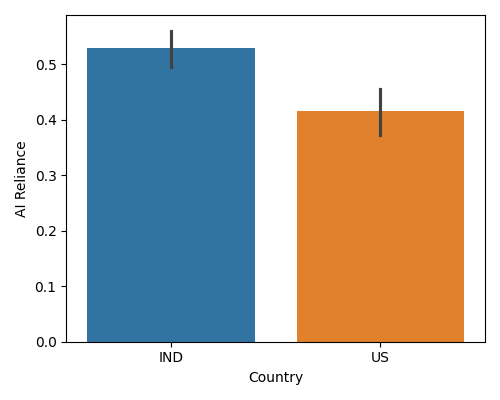} & \includegraphics[height=1.47in]{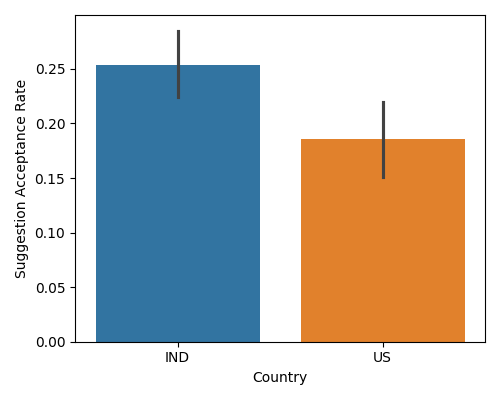} &
    \includegraphics[height=1.47in]{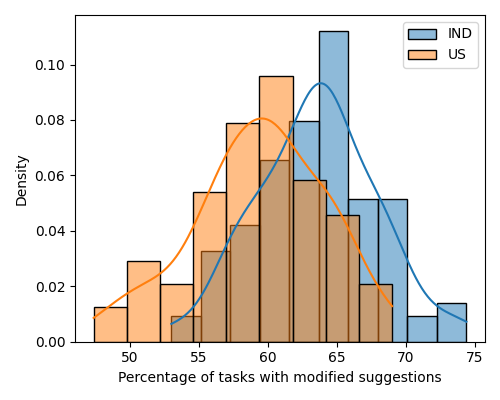} \\
    (a) AI reliance & (b) Suggestion acceptance rate & (c) Suggestions modified
    \end{tabular}
    \caption{Various metrics of engagement with AI suggestions. Indian participants show a higher engagement with AI suggestions but also need to modify them more often.}
    \Description{Three bar graphs labeled (a), (b), and (c). Graph (a) shows "AI reliance" for Indian (IND) and US participants, with Indians having higher AI reliance. Graph (b) shows the "Suggestion acceptance rate" for the two cohorts, with Indians accepting more suggestions. Graph (c) is a histogram displaying the "Percentage of tasks with modified suggestions." Distributions for India and the US overlap, but the distribution for India is shifted to the right.}
    \label{fig:ai_engagement}
\end{figure*}

\subsubsection{Engagement with AI Suggestions}
We measure engagement using two metrics defined previously: AI reliance and suggestion acceptance rate. In short, AI reliance measures the proportion of an essay written using AI suggestions, while the acceptance rate measures the proportion of suggestions the user accepts in a task. Section~\ref{subsec:analysis} provides detailed descriptions of these metrics.

Figure~\ref{fig:ai_engagement}(a) illustrates the AI reliance scores for essays written by Indian and American participants. We observe that Indians exhibited a higher reliance on AI than Americans. The mean AI reliance score for Indians was $0.53$ ($SD=0.2$), suggesting that approximately half of their essays were AI-written, while the average for Americans was $0.42$ ($SD=0.25$). The Mann-Whitney U test revealed that this difference was statistically significant ($U = 10614.0$, $p < 0.001$) with a small effect size (Cliff's delta $D = 0.27$).

Figure~\ref{fig:ai_engagement}(b) shows a similar plot for the suggestion acceptance rate. Again, on average, Indians accepted a higher proportion of the suggestions shown to them in a task. On average, Indian participants accepted $25\%$ ($SD=19$) of the suggestions in a task, while American participants accepted $19\%$ ($SD=19$). The Mann-Whitney U test confirmed that this difference was statistically significant ($U = 10748.0$, $p < 0.001$) with a small effect size (Cliff's $D = 0.29$).

Overall, these results indicate that \textbf{Indians showed a higher engagement with AI suggestions than Americans on both metrics. This disparity might stem from differences in how AI is perceived or used in different cultural contexts.} For example, non-native English speakers might find in-situ AI suggestions more appealing than native speakers, as shown by \citet{Buschek2021}.

\subsubsection{Suggestion Modification Behavior}
Users may accept a suggestion and then modify it to better suit their context. Since the suggestions were Western-biased (see Table~\ref{tab:top_suggestions}), modification may be required to make them culturally relevant for Indians. To analyze the modification behavior, we compared the percentage of tasks in which Indians and Americans modified at least one suggestion. 

Figure~\ref{fig:ai_engagement}(c) shows the distribution of the proportion of tasks where Indians and Americans modified at least one suggestion. To produce each distribution, we conducted a bootstrap sampling procedure over 100 iterations. In each iteration, we sampled 80\% of the tasks (with replacement) and calculated the percentage of tasks in which Indians and Americans modified at least one suggestion. This produced a distribution of percentages (one for each of the 100 bootstrap iterations) for Indians and Americans, which we visualized using a density plot.

We observe that the distribution for Indian participants (blue curve) is shifted to the right. On average, Indian participants modified suggestions in $63.5\%$ of the tasks ($SD=4.2$) whereas American participants modified suggestions in $59.4\%$ of the tasks ($SD=4.8$).
A t-test revealed that this difference was significant, $t(198) = 6.48$, $p < 0.001$, with a large effect size (Cohen's d was $0.91$).

These results suggest that \textbf{Indian participants modified AI-generated suggestions in significantly more tasks than American participants}, indicating that the suggestions were less suitable for Indian participants in their original form. Taken together with the engagement metrics presented above, this shows that \textbf{although Indians accepted more suggestions, they also had to modify them more frequently to better fit their writing style and context}. Next, we investigate the impact of this heightened cognitive load on the productivity of Indian participants.

\subsubsection{Productivity Derived from AI Suggestions}

\begin{figure*}[t]
    \centering
    \begin{tabular}{ccc}
    \includegraphics[height=1.47in]{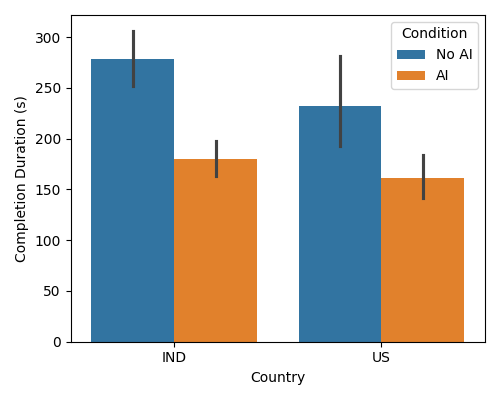} & \includegraphics[height=1.47in]{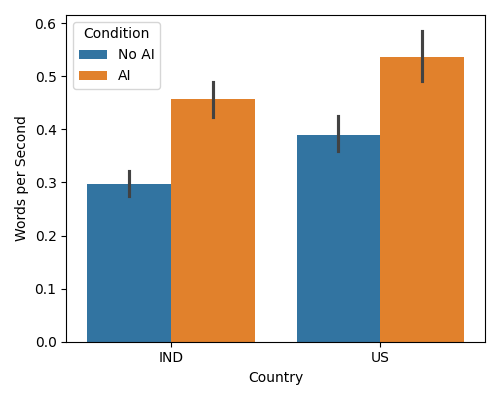} & \includegraphics[height=1.47in]{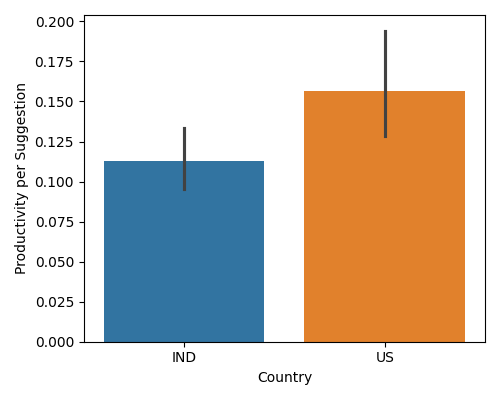} \\
    (a) Completion duration & (b) Productivity (words per second) & (c) Productivity per suggestion
    \end{tabular}
    \caption{Productivity measures. (a, b) Both cohorts experience productivity gains from writing with AI suggestions. (c) However, the productivity derived per suggestion is lower for Indian participants.}
    \Description{Three bar graphs labeled (a), (b), and (c). Graph (a) shows "Completion duration" in seconds for Indian (IND) and US participants, with blue bars representing the "No AI" condition and orange bars representing the "AI" condition. Both cohorts show shorter completion times with AI, with Indian participants taking longer overall. Graph (b) displays "Productivity (words per second)" under the same structure. Both cohorts show higher higher productivity with AI. US participants achieve higher productivity rates than Indian participants in both conditions. Graph (c) shows "Value derived per suggestion." It has two bars representing the two countries. US participants derive more value per AI suggestion than Indian participants.}
    \label{fig:productivity_metrics}
\end{figure*}

Next, we compare the productivity gains derived from AI suggestions by Indian and American participants, focusing on the task completion time and the productivity measure defined in Section~\ref{subsec:analysis}.

Figure~\ref{fig:productivity_metrics}(a) illustrates the task completion duration for both Indian and American participants. \textbf{Both cohorts showed a reduction in completion times when AI suggestions were used.} For Indians, the mean task completion time decreased from $278.4$ seconds without AI ($SD=134.5$) to $179.6$ seconds with AI ($SD=105.8$), representing a 35\% decrease ($\sim$1.5 minutes faster). This decrease was statistically significant ($U=10356.0$, $p < 0.001$) with a large effect size (Cliff's $D = 0.5$). Similarly, Americans experienced a decrease, from $232.0$ seconds without AI (SD=$240.1$) to $161.1$ seconds with AI ($SD=117.7$), a 30\% decrease ($\sim$1.2 minutes faster). This reduction was also significant ($U=8936.0$, $p<0.001$) with a medium effect size (Cliff's $D = 0.33$). While descriptively it seemed that Indians gained slightly more from AI (35\% decrease vs 30\% for Americans), a Difference-in-Difference analysis using an Ordinary Least Squares (OLS) regression model showed that the reduction in duration was similar in both cohorts ($p=0.34$, 95\% CI=$[-85.25, 29.58]$).

Nevertheless, the descriptive difference in reduction prompted us to further investigate potential productivity gains. Figure~\ref{fig:productivity_metrics}(b) presents the productivity of participants (words written per second) in all four groups. \textbf{Both Indian and American participants experienced a significant productivity boost when using AI suggestions} ($p < 0.001$ for both cohorts, measured by the Mann-Whitney U test). However, the effect size was larger for Indians (Cliff's $D = -0.48$ compared to $D = -0.33$ for Americans), suggesting a slightly higher productivity gain for Indians. However, as shown previously, this higher gain came at the back of a higher suggestion acceptance rate. This raised the question: Was the higher productivity substantial enough to justify the higher acceptance rate?

To answer this, we calculated the productivity derived \textit{per suggestion} by normalizing the productivity by the number of suggestions accepted\footnote{This metric is only applicable to the AI condition, as suggestions were not provided in the control condition.}. However, this metric may understate the productivity derived by Indians, as they accepted more suggestions, and suggestions tend to offer diminishing returns (each additional suggestion provides less incremental value). To address this, we limited this analysis to essays where participants accepted up to seven suggestions\footnote{Very few Americans accept more than seven suggestions in a task, whereas Indians continue to accept more, making comparisons unreliable.}. Figure~\ref{fig:productivity_metrics}(c) confirms our hypothesis that \textbf{Indians derived significantly less productivity per suggestion compared to Americans} ($U=5424.0$, $p < 0.001$, $D=-0.27$).


Overall, these results indicate that \textbf{AI helped both Indian and American participants write faster, but Americans derived more productivity from each suggestion, leaving Indian participants to rely more heavily on AI suggestions to achieve similar productivity outcomes}. This may be due to Indians overrelying on AI beyond the point that it was helpful, or, as concluded from their suggestion modification behavior because the AI suggestions were inherently less effective for them (i.e., higher cognitive load required to modify AI suggestions compared to American participants).

%% file: 4.2_.tex
\subsection{Does AI homogenize writing within cultures?}

In the previous section, we outlined how AI impacts Indian and American users differently. In this section, we show that AI homogenizes writing styles within the same culture. This will set the stage for our main investigation of cross-cultural homogenization in the next section.

\subsubsection{Similarity Analysis} \label{subsub:intra_similarity}
We conducted a similarity analysis to investigate whether AI standardizes writing styles for participants within their respective cultures. The similarity metric we used was cosine similarity, computed on the essay embeddings obtained from OpenAI's text-embedding model.

\textbf{Setup:} For this experiment, we computed similarity scores within each of our four experimental groups and compared the four distributions. For example, there were $n=24$ Indian participants in the No AI condition, and each participant wrote four essays. So, we computed pair-wise similarity scores between all essays written by these 24 participants, grouped by the essay topic. This yielded a distribution of $4 \text{ essay topics } \times \binom{24}{2} \text{ comparisons per topic}  = 1104$ similarity scores. Similarly, we computed similarity distributions for the other three experimental groups (Indians with AI, Americans without AI, and Americans with AI). The means and 95\% confidence intervals of these distributions are shown in Figure~\ref{fig:similarity_scores}(a). The blue line shows how the within-culture similarity among Indian participants changes from the No AI condition to the AI condition, while the orange line represents this change for American participants.

\textbf{Results:} First, we notice that without AI (left side of the graph), Indian participants exhibited slightly higher within-culture similarity than American participants, suggesting that Indians naturally had more homogeneous writing styles than Americans. This natural difference between the Indian and American cohorts was significant, $t(2726) = 4.04$, $p < 0.001$, though the effect size was negligible (Cohen's $d = 0.16$).

Next, when AI suggestions were introduced, the similarity increased within both Indian and American cohorts. A t-test revealed that this increase was significant within both cohorts; $t(3622)=-7.22$, $p < 0.001$ for Indians (blue line) with a small effect size (Cohen's $d = -0.26$), and $t(3246)=-9.45$, $p < 0.001$ for Americans (orange line) with a small effect size (Cohen's $d = -0.33$). This means that AI caused both Indians and Americans to write more similarly within their cohorts. Visually, the magnitude of the increase looks slightly higher for Americans (the orange line is steeper than the blue line). However, a Difference-in-Difference analysis using a regression model showed that the increase was statistically similar in both cohorts ($p=0.30$, 95\% CI=$[-0.018, 0.006]$). This means that the magnitude of within-culture homogenization was similar in both cohorts.

\textbf{In sum, the increasing similarity in both cohorts suggests that AI made writing more homogeneous, potentially by guiding users towards common phrasing, structure, or content.} Thus, AI suggestions had a net homogenizing effect, suggesting that some level of cross-cultural homogenization was to be expected. However, as we demonstrate next, the extent of this homogenization was even more pronounced across cultures.

%% file: 4.3_.tex
\subsection{Does AI homogenize writing across cultures?}
We also investigated if AI might homogenize writing styles across cultures, i.e., do Indians start writing like Americans (or vice versa) when they use AI suggestions? We now show evidence for this effect.

\subsubsection{Similarity Analysis}

\begin{figure*}[t]
    \centering
    \begin{tabular}{ccc}
    \includegraphics[height=1.46in]{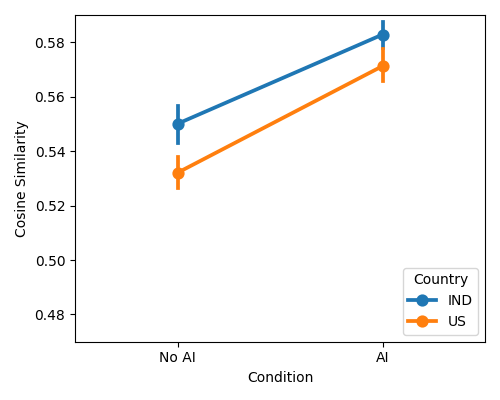} & 
    \includegraphics[height=1.46in]{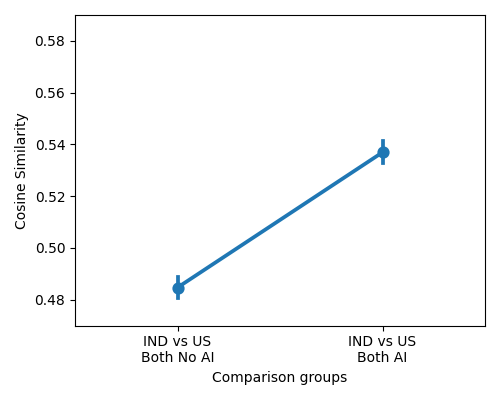} & \includegraphics[height=1.46in]{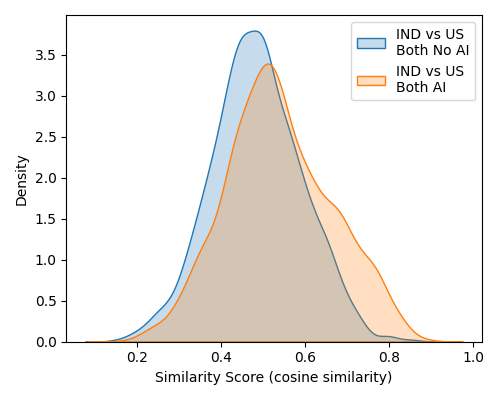} \\
    (a) Within-culture similarity & (b) Cross-culture similarity & (c) Cross-culture similarity
    \end{tabular}
    \caption{Within- and cross-culture similarity between Indian and American essays when written without and with AI. (a) The within-culture similarity is higher when AI is used, suggesting that AI homogenizes writing within both cohorts. (b) The cross-culture similarity is higher when AI is used, showing that AI homogenizes writing styles across cultures. (b) Cross-culture similarity scores are higher when AI suggestions are used.}
    \Description{Three graphs labeled (a), (b), and (c). Graph (a) shows "Within-culture similarity" (cosine similarity) for Indian (IND) and US participants under "No AI" and "AI" conditions. Both cohorts show increased similarity with AI, with Indians displaying slightly higher overall within-culture similarity both without and with AI. Graph (b) shows "Cross-culture similarity" between Indian and US participants. The similarity score increases significantly when both cohorts use AI suggestions, indicating more homogeneous writing styles with AI. Graph (c) is a density plot of the "Cross-culture similarity" scores (same data as Graph b, but shown as distributions). The distribution for the AI condition shifts right, showing higher similarity scores when AI suggestions are used.}
    \label{fig:similarity_scores}
\end{figure*}

In this experiment, we extend the previous similarity analysis that demonstrated within-group homogenization by performing a cross-cultural comparison of essay embeddings.

\textbf{Setup:} First, we calculated similarity scores between essays written by Indians and Americans without AI, grouped by topic. There were $n=24$ Indians and $m=29$ Americans in the No AI condition, and each of them wrote essays on four topics. Hence, we computed the pairwise cosine similarity between all essays for each topic, yielding a distribution of $4 \text{ essay topics} \times (24 \times 29) \text{ comparisons per topic} = 2784$ cross-culture similarity scores. This distribution represented the similarity between Indian and American essays when AI suggestions were not used. Next, we computed a distribution of similarity scores when AI suggestions \textit{were} used. Figure~\ref{fig:similarity_scores}(c) shows these two distributions.

\textbf{Results:} Figure~\ref{fig:similarity_scores}(b) compares the means and 95\% confidence intervals of the two distributions. It shows that without AI, the mean similarity between Indians and Americans was $0.48$ $(SD=0.11)$, but with AI, the similarity score rose to $0.54$ ($SD=0.13$). A t-test revealed that this increase was statistically significant, $t(6958)=-17.89$, $p < 0.001$. This indicates that on average, Indians and Americans wrote more similarly when AI suggestions were used. Further, the effect size of this increase (Cohen's $d = -0.44$) is larger than what was observed in the within-culture homogenization for both cohorts, indicating a stronger cross-cultural homogenization effect.

Figure~\ref{fig:similarity_scores}(c) confirms this observation. The similarity distribution in the AI condition (orange curve) is shifted to the right compared to the No AI condition (blue curve), showing a skew toward higher similarity scores. Moreover, the AI condition shows a longer right tail, extending further into higher similarity scores than the No AI distribution. This confirms that Indian and American participants wrote more similar essays when AI suggestions were used. Overall, these results suggest that \textbf{AI influenced Indian and American participants to write more similarly, thereby diminishing cultural differences in writing}.

\bheading{Direction of Homogenization}
The above analysis reveals cross-cultural homogenization. However, since cosine similarity is a symmetric measure, it does not indicate the direction of influence---whether Indians adopt American writing styles or vice versa. To investigate this, we conducted a complementary test. We computed similarity scores when only one of the two cohorts was using AI: (a) between Americans not using AI and Indians using AI, and (b) vice versa, Indians not using AI and Americans using AI. This would allow us to see whether AI is making Indian writing styles converge more strongly toward natural American styles, or vice versa.

In both scenarios, we observed significantly higher similarity scores when one cohort used AI, compared to when neither cohort used AI ($p < 0.001$, based on independent t-tests). This is consistent with our earlier finding that AI has a net homogenizing effect. However, effect sizes revealed an important distinction: AI caused Indians to write more similarly to Americans' natural styles (Cohen's $d=-0.22$; small) than the other way around (Cohen's $d=-0.12$; negligible). \textbf{This suggests that AI may be driving Indian writing toward Western styles more than it is influencing American writing toward Indian styles.} In contrast, the reverse influence (Americans adopting Indian writing styles) was less pronounced. This finding is supported by prior research showing that AI models exhibit Western biases~\cite{johnson2022ghost, cao2023chatgptculture}, which may be causing a stronger convergence of Indian writing toward Western styles.


\subsubsection{Predictive Modeling} \label{subsubsec:predictive_modeling}

\begin{figure*}[t]
    \centering
    \begin{tabular}{c}
    \includegraphics[width=0.95\textwidth]{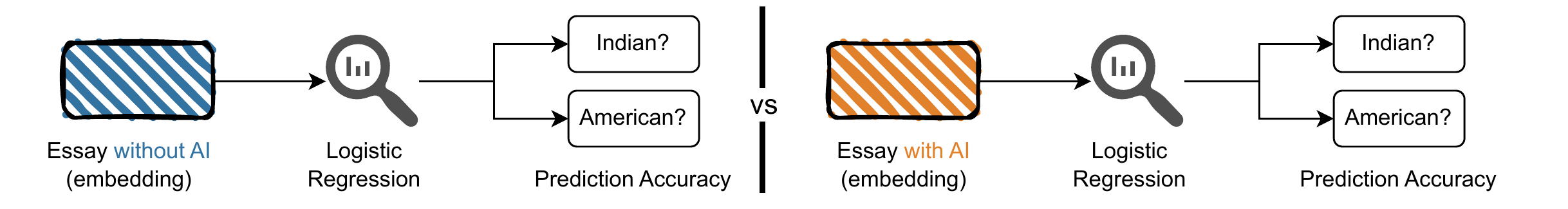}
    \end{tabular}
    \caption{Predictive modeling to investigate homogenization across cultures. Full details in Section~\ref{subsubsec:predictive_modeling}.}
    \label{fig:predictive_modeling}
    \Description{A diagram showing two predictive modeling processes. On the left, a block labeled "Essay without AI (embedding)" leads to "Logistic Regression," which predicts whether the writer is "Indian" or "American". This classification is shown to have a prediction accuracy. On the right, a similar flow is shown for "Essay with AI (embedding)." The diagram highlights the comparison between prediction accuracy in the "without AI" and "with AI" conditions.}
\end{figure*}

We now demonstrate the homogenization effect using another experiment. We investigated whether AI suggestions make it harder for a classification model to differentiate Indian vs American authorship. If true, it would indicate that AI was diminishing cultural differences in writing styles.

\textbf{Setup:} We trained a classification model to determine whether an essay was written by an Indian or American participant. For simplicity and efficiency, we chose a logistic regression model, which is particularly well-suited for binary classification tasks. As shown in Figure~\ref{fig:predictive_modeling}, we used essay embeddings obtained from OpenAI's text-embedding model as features, with the participant's cohort (Indian or American) as the label. To isolate the impact of AI, we compared the performance of two models: one trained on essays written without AI suggestions (No AI condition) and the other on essays written with AI suggestions (AI condition). To ensure the robustness of our experiment, we used stratified k-fold cross-validation with $k=20$, building 20 different models for each condition by swapping out one of the 20 folds as the test set. This resulted in a distribution of 20 accuracy and F1 scores in each condition, which we compared using an independent samples t-test.

\textbf{Results:} Our results show that the accuracy and F1 scores were lower in the AI condition. On average, the accuracy decreased by $7.1$ percentage points, from $90.6\%$ in the No AI condition to $83.5\%$ in the AI condition. This decrease was statistically significant, $t(38) = 2.25$, $p = 0.03$, with a medium effect size (Cohen's $d = 0.71$). Similarly, the macro F1 score decreased by $7$ percentage points on average, from $89.9\%$ in the No AI condition to $82.9\%$ in the AI condition, $t(38) = 2.07$, $p = 0.045$, also with a medium effect size (Cohen's $d = 0.65$).

\textbf{The significant reduction in both accuracy and F1 scores indicates that a classification model found it more challenging to differentiate between Indian and American authorship when AI suggestions were used.} This provides further evidence that AI diminished the natural writing differences between Indian and American participants, potentially steering users toward more Western-centric writing styles.

\subsubsection{Did AI cause deep writing changes or merely omit explicit cultural references?}

\begin{figure*}[t]
    \centering
    \begin{tabular}{ccc}
    \includegraphics[height=1.46in]{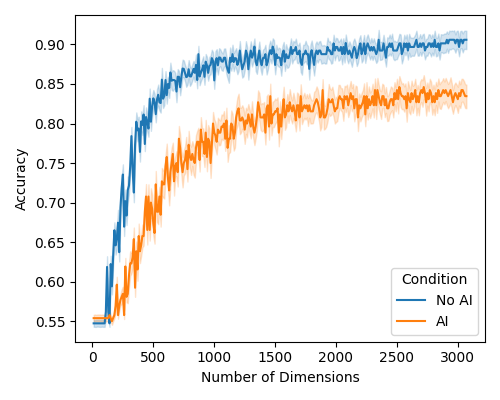} & \includegraphics[height=1.46in]{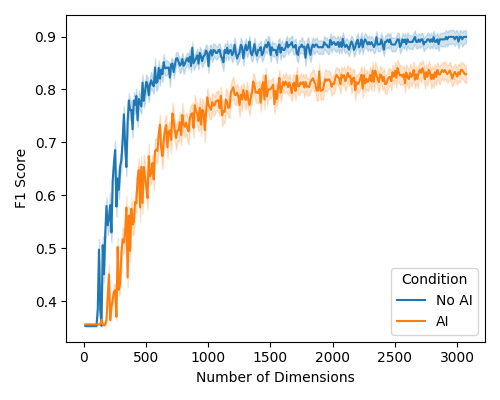} & \includegraphics[height=1.46in]{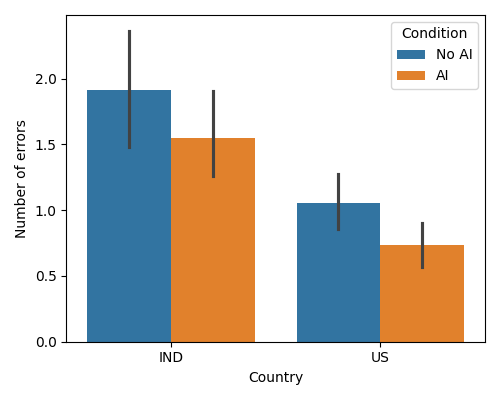} \\
    (a) Accuracy & (b) F1 score & (c) Grammatical errors
    \end{tabular}
    \caption{(a, b) A classification model trained to predict Indian vs American authorship consistently performs worse on essays written with AI suggestions. (c) AI reduces grammatical errors in both cohorts, showing that the homogenization effect is not due to mere grammatical fixes.}
    \Description{Three graphs labeled (a), (b), and (c). Graph (a) shows the "Accuracy" (y-axis) of a classification model predicting Indian vs. American authorship for sub-embedding sizes (the x-axis is the number of dimensions). The blue line represents the "No AI" condition, while the orange line represents the "AI" condition. Accuracy is consistently higher without AI suggestions. Graph (b) shows the same graph but for "F1 Score" on the y-axis. Again, the "No AI" condition (blue) performs better than the "AI" condition (orange), with the gap remaining consistent across all dimension sizes. Graph (c) is a bar chart comparing the "Number of grammatical errors" for Indian (IND) and US participants. AI reduces grammatical errors for both cohorts, with Indian participants showing more errors overall in both conditions.}
    \label{fig:dimension_grammar_analysis}
\end{figure*}

Three of the four writing tasks in our study prompted participants to describe explicit cultural artifacts, such as food, public figures, and festivals. These explicit references can make it easier for a classification model to identify the writer's culture. So, if AI suggestions steered Indian participants away from these cultural markers, the absence of such references could explain the model's reduced performance. However, we wanted to explore whether AI also caused more subtle changes in writing style that went beyond the mere omission of explicit cultural artifacts (e.g., uniquely Indian food items, public figures, and festivals). We conducted two additional analyses to investigate this.

\textbf{Ablation on Embedding Dimensions:} We examined whether the performance drop persisted when the dimensionality of the essay embeddings was reduced. Specifically, we randomly selected $d$ dimensions ($d < 3072$) from the original 3072-length embeddings and conducted the classification task using only these sub-sampled embeddings as features. To test this, we repeated the k-fold cross-validation experiment ($k=20$) described earlier but varied the length of the essay embeddings from $d=12$ to $d=3072$ in increments of 10. In each iteration, we randomly sampled $d$ dimensions from the original embedding and used only those dimensions as features for the model. To ensure robustness to random sampling, we repeated the experiment 10 times for each value of $d$.

The results, shown in Figure~\ref{fig:dimension_grammar_analysis}(a, b), reveal a consistent performance drop in both accuracy and F1 score across all embedding sizes in the AI condition. Importantly, this drop persists even for embeddings with very low dimensionality, which are unlikely to capture explicit cultural artifacts (e.g., uniquely Indian food items or public figures). By randomly sampling dimensions, we disrupted the structure of the embedding space, further reducing the likelihood that specific cultural features would be preserved. Despite this, the performance gap suggests that the reduced accuracy is not merely due to the absence of these cultural references, but reflects deeper changes in writing style introduced by AI.

\textbf{Performance on Email-Writing Task:} Next, we trained the model using only the essays from the email writing task and examined the resulting accuracy and F1 scores. This task (requesting a leave of absence) was designed to capture implicit cultural nuances rather than explicit references. We found that the model's performance in distinguishing between Indian and American authorship in the AI condition was even worse for these essays. Accuracy dropped from $82.9\%$ in the No AI condition to $60\%$ in the AI condition, a statistically significant decrease, $t(8)=5.63$, $p < 0.001$, with a large effect size (Cohen's $d = 3.56$). The F1 score also declined sharply, from $81.7\%$ to $49.2\%$ AI, $t(8)=5.79$, $p < 0.001$, with a large effect size ($d=3.66$). This supports the findings that AI suggestions may be influencing subtle aspects of writing style, contributing to a homogenization effect that goes beyond the mere omission of explicit cultural identifiers.

\textbf{These two experiments suggest that the decreased model performance was not merely due to the absence of explicit cultural markers, instead it reflects deeper, more subtle changes in writing style induced by AI.} This is a crucial finding, as it indicates that \textbf{AI suggestions influence not just the surface content of what is written, but also more ingrained elements of \textit{how} it is written}.

\subsubsection{Did AI homogenize Indian writing solely by correcting grammar?}
Another plausible explanation for the homogenization effect could be that, as non-native English speakers, Indian writers made more grammatical errors, and AI suggestions corrected these errors, leading to the observed homogenization. To test this hypothesis, we ran a grammar check on all essays using an open-source tool called LanguageTool\footnote{\href{https://languagetool.org/}{LanguageTool}, used through a Python library: \texttt{\href{https://github.com/jxmorris12/language_tool_python}{language\_tool\_python}}.}. This tool applies over 6000 grammar rules to the text. However, its spell-checker misclassifies Indian proper nouns as grammatical errors, overstating errors for Indian essays. To ensure a fair comparison, we disabled the spell-checking rules while keeping all other grammar rules intact.

The results of this analysis are shown in Figure~\ref{fig:dimension_grammar_analysis}(c). Descriptively, it can be seen that AI reduced grammatical errors in both Indian and American writing, but the reduction was similar in both cohorts. Statistically, the reduction was significant in American essays ($U=7729.0$, $p=0.04$) with a small effect size (Cliff's Delta $D=0.15$), but \textit{not} significant in Indian essays ($U=7584.5$, $p=0.18$).

If grammar correction was the primary driver of homogenization, we would expect a larger reduction in grammatical errors for Indian writers, given their higher propensity of making such errors. However, we see that AI's impact on grammar correction was similar for both Indians and Americans, \textbf{the homogenization effect cannot be attributed solely to grammatical corrections}. Instead, AI appears to be influencing writing in more subtle ways, which we illustrate next with a concrete example.

\subsubsection{A Concrete Example: Altering Lexical Diversity}

\begin{figure*}[t]
    \centering
    \begin{tabular}{cc}
    \includegraphics[height=1.48in]{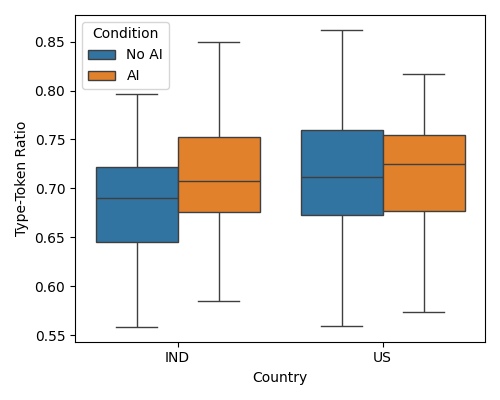} & \includegraphics[height=1.48in]{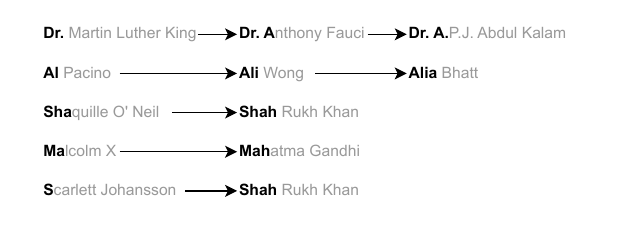} \\
    (a) Lexical diversity (TTR) & (b) Culturally incongruent suggestions\\
    \end{tabular}
    \caption{(a) Lexical diversity of the essays (measured by type-token ratio) partitioned by country and AI conditions. On average, AI increases the lexical diversity for Indians, taking it closer to American styles, but does not change the lexical diversity for Americans. Note: We show a box plot because, for this analysis, the bar plot under-emphasized statistically significant differences. (b) Progression of AI suggestions for public figures. Participants typed black text to guide suggestions, but the system initially proposed culturally incongruent options (gray). The intended match emerged only after adding more context.}
    \Description{Two subfigures labeled (a) and (b). Subfigure (a) is a box plot showing "Lexical diversity (TTR)" for Indian (IND) and US participants under "No AI" (blue) and "AI" (orange) conditions. For Indian participants, AI increases lexical diversity (type-token ratio), bringing it closer to the American participants' level. For US participants, the TTR stays somewhat similar in No AI and AI conditions. Subfigure (b) depicts the progression of AI autocomplete suggestions for public figures. Suggestions are initially culturally incongruent, such as "Dr. Martin Luther King" for Indian users, but are replaced with culturally appropriate suggestions like "Dr. A.P.J. Abdul Kalam" when more context is provided.}
    \label{fig:ttr}
\end{figure*}

Our two main experiments (cosine similarity and predictive modeling) used word embeddings to capture the semantic meaning of participants' essays, providing a robust yet abstract measure of homogenization. We also showed that the homogenization effect is not simply the result of omitting cultural references or correcting grammatical errors, meaning that AI induced more subtle, implicit changes in writing. What are some of these changes?

To provide a concrete example, we examined the lexical diversity of essays written with and without AI. We used the Type-Token Ratio (TTR), a commonly used metric to measure lexical diversity. A higher TTR reflects a greater usage of unique words, indicating higher lexical diversity.

Figure~\ref{fig:ttr}(a) shows the lexical diversity in participants' essays, partitioned by country and AI condition.
The natural lexical diversity in Indian writing (without AI) was significantly different from that of American participants (the two blue boxes in the plot), $t(210) = -2.94$, $p = 0.003$, Cohen's $d = -0.41$. However, when AI suggestions were introduced, the lexical diversity of Indians went up and converged with that of Americans (the two orange boxes) and the difference between the two cohorts was no longer significant, $t(258) = -0.32$, $p = 0.75$, Cohen's $d = -0.04$. Thus, AI eliminated the significant difference that existed between their natural styles. 

This suggests that \textbf{AI homogenized the natural lexical diversity exhibited by Indian participants, aligning it more closely with the diversity typical of American writing styles}. While our initial analyses reveal homogenization using high-dimensional text embeddings, the converging TTR provides a concrete example of homogenization toward Western writing styles.


%% file: 4.4_.tex
\subsection{Impact of AI on Cultural Writing}
We conducted a content analysis to examine the writing changes introduced by AI. As part of this process, we manually coded essays on the first three topics (favorite food, festival, and public figure) to identify references to cultural symbols and idols. We focused on essays written by Indian participants, as the aforementioned quantitative results indicated that homogenization had a greater impact on them compared to the American participants.

\subsubsection{Changing Cultural Expression}
\input{tables/essays}
To identify how AI induced changes in the cultural expression of Indian participants, we compared essays about the same festivals and food items written with and without AI, focusing on dimensions such as exoticization, misrepresentation, and Western gaze~\cite{Qadri2023regimes}. Table~\ref{tab:essay_examples} provides illustrative examples highlighting how AI introduced changes in cultural expression and more examples are provided in Appendix~\ref{appendix:essay_examples_appendix}.

As shown in Table~\ref{tab:essay_examples} essays describing the Indian festival of Diwali without AI often included rich cultural details such as religious context and worship rituals. In contrast, essays written with AI tended to feature more generic descriptions, emphasizing elements like sweets, gift exchanges, and family gatherings. While these descriptions are not incorrect and do not constitute overt misrepresentation, they lack cultural nuance. For example, family gatherings during Diwali often involve specific activities such as designing rangolis and bursting firecrackers. These details appeared in essays without AI but were absent in those written with AI, which reduced Indian festivals to generic celebrations.
Similarly, when writing about Biryani, participants in the No AI condition described specific regional variations and accompaniments. However, essays written with AI did not contain such details and frequently used cultural tropes like ``rich'', ``flavorful'', and ``aromatic'', subtly exoticizing Indian food. This analysis was inconclusive for the public figures task due to the vast pool of celebrities, resulting in a minimal overlap between those mentioned in the AI and No AI conditions.

\subsubsection{Cultural Incongruence in AI Suggestions}
Given the open-ended nature of the tasks, we could not computationally record the exact moment an artifact was suggested to a participant. Hence, we manually analyzed the logs, examining the suggestions offered by the AI system and the text preceding them. Starting with the public figure task, we recorded the sequence of celebrities suggested by the system as participants added more characters until they finalized the name in their essay. Each suggested celebrity was then categorized as Indian or non-Indian based on their citizenship.

Among Indian participants, we found no instance where the first suggested celebrity matched the one ultimately included in the essay. Moreover, the initial suggestions were almost exclusively Western figures (with one exception). This resulted in frequent mismatches, where culturally incongruent suggestions persisted even after participants began typing their intended public figure’s name. Examples of such mismatches are shown in Figure~\ref{fig:ttr}(b).

The food and festival tasks showed fewer stark mismatches due to a smaller set of possible suggestions for similar starting characters. However, the first suggested food item was always pizza or sushi, and the first suggested festival was invariably Christmas. We build on these observations next.


\subsubsection{Did AI suggestions influence participants' choice of artifacts?}
Our study design did not allow us to capture what participants in the AI condition \textit{would} have written \textit{without} AI suggestions. However, we compared the artifacts selected by participants across the AI and No AI conditions.

In the food task, the first suggested item was always pizza or sushi. While pizza is popular in India, no Indian participant selected sushi as their favorite dish in the No AI condition, yet sushi appeared in three essays in the AI condition. Similarly, the first AI suggestion in the festival task was always Christmas. Without AI suggestions, 12.5\% of Indian participants (3/24) wrote about Christmas, but this proportion slightly increased to 13.9\% (5/36) in the AI condition. In the public figures task, 54\% of Indian participants (13/24) selected Indian public figures in the No AI condition, compared to 44\% (16/36) in the AI condition (full list in Appendix~\ref{appendix:celebrities_selected}).

These observations suggest a possibility of changing cultural preferences due to the Western bias of the model. However, the small sample size limits the strength of our conclusions, especially since participants often bypassed the first suggestions in favor of their actual preferences. This trend merits further investigation in future research.


%% file: tables/essays.tex
\begin{table*}[t]
\small
\centering
\begin{tabular}{lll}
\multicolumn{1}{c|}{\textbf{Topic}} & \multicolumn{1}{c}{\textbf{Essay without AI}} & \multicolumn{1}{c}{\textbf{Essay with AI}} \\ \hline
\multicolumn{1}{l|}{\begin{tabular}[c]{@{}l@{}}Festival\\ (Diwali)\end{tabular}} & \begin{tabular}[c]{@{}l@{}}My favorite festival is Diwali. Diwali is usually\\ celebrated in between October and November.\\ On this occasion, \colorbox{PastelGreen}{I used to worship goddess}\\ \colorbox{PastelGreen}{Laxmi} along with my family. Most fascinating\\ thing is to pop crackers and eat sweets for\\ minimum span of 4 days. I always wait for this\\ festival. We lighten our houses with earthen\\ lamps and \colorbox{PastelGreen}{worship cows and lords}.\end{tabular} & \begin{tabular}[c]{@{}l@{}}My favorite festival is Di\textbf{wali because of}\\ \textbf{the }\colorbox{PastelYellow}{\textbf{lights and sweets}}. Its the bigg\textbf{est}\\ \textbf{festival in India}. In morning we eat\\ \colorbox{PastelBlue}{\textbf{traditional Indian breakfast items}}. We\\ also clean or pain\textbf{t our homes to welcome}\\ \textbf{prosperity}. At night we \textbf{light diyas and}\\ \textbf{burst fireworks}. Its \textbf{a time for family}\\ \textbf{gatherings and }\colorbox{PastelYellow}{\textbf{joyous celebrations}}.\\ Everyone \textbf{wears new clothes and}\\ \colorbox{PastelYellow}{\textbf{exchanges gifts}}. Its \textbf{a time filled with}\\ \colorbox{PastelYellow}{\textbf{happiness and warmth}}.\end{tabular} \\ \hline
\multicolumn{1}{l|}{\begin{tabular}[c]{@{}l@{}}Food\\ (Biryani)\end{tabular}} & \begin{tabular}[c]{@{}l@{}}My favorite food is chicken biriyani. I like it\\ because it tastes good and it is easy to prepare.\\ I like to prepare it in \colorbox{PastelGreen}{Malabar style}. That recipe\\ uses some exotic \colorbox{PastelGreen}{ingredients like nutmeg}.\\ Chicken biriyani is \colorbox{PastelGreen}{along with raita, lemon}\\ \colorbox{PastelGreen}{pickle and dates chutney} tastes divine. Chicken\\ biriyani is supposed to be a dish that was\\ brought to India by Mughals.\end{tabular} & \begin{tabular}[c]{@{}l@{}}my favorite food is briyani because \textbf{of its}\\ \colorbox{PastelOrange}{\textbf{rich flavors and spices}}. i love  \textbf{the way it}\\ \colorbox{PastelOrange}{\textbf{melts in my mouth}}. i use  \textbf{aromatic}\\ \textbf{basmati rice and tender meat}. i love\\ chicken \textbf{biryani the most}. which \\ \textbf{is your favorite type of biryani?}\\ hyderbadi \textbf{biryani is also delicious}. which\\ has \colorbox{PastelOrange}{\textbf{rich spices} and a unique taste}.\end{tabular} \\
 &  & 
\end{tabular}
\caption{Comparison of cultural artifacts described by Indian participants with and without AI suggestions. AI-generated text is shown in \textbf{bold}. Highlights indicate key differences: \colorbox{PastelGreen}{green} for cultural details present without AI but missing with AI, \colorbox{PastelYellow}{yellow} for generic phrasing introduced by AI, \colorbox{PastelBlue}{blue} for Westernized descriptions, \colorbox{PastelOrange}{orange} for exoticization of Indian food.}
\Description{A table comparing cultural artifacts described in essays by Indian participants with and without AI suggestions. Two topics are shown: "Festival (Diwali)" and "Food (Biryani)." For "Festival (Diwali)," the essay without AI mentions traditional worship practices such as worshipping Goddess Laxmi and cows, while the essay with AI highlights generic descriptions like "lights and sweets," "family gatherings," and "joyous celebrations." For "Food (Biryani)," the essay without AI discusses preparation methods (e.g., "Malabar style" and "ingredients like nutmeg") and accompaniments (e.g., "lemon pickle and dates chutney"), while the essay with AI emphasizes rich and flavorful descriptions like "rich flavors and spices" and "melts in my mouth."}
\label{tab:essay_examples}
\end{table*}

%% file: 5_discussion.tex
Our results show that AI-powered writing suggestions yield greater benefits for American users compared to Indian users. Moreover, AI suggestions homogenize writing styles between Indians and Americans, pushing Indian users to adopt American writing styles and diminishing cultural nuances in writing. Worryingly, the changes induced by AI suggestions are both explicit (e.g., changing cultural preferences), as well as subtle and deeply ingrained (e.g., lexical diversity, exoticization, Westernization), making it harder to identify the cultural erasure they might be causing. In the following sections, we discuss cultural differences in AI attitudes, examine the implications of the observed homogenization of writing styles, and propose strategies to mitigate the harms of cultural homogenization and AI-driven neocolonialism.

\subsection{Cultural Differences in AI Attitudes}
As we showed in Section~\ref{subsec:differential_impact}, Indian participants exhibited higher engagement with AI suggestions: a larger proportion of their essays were composed of AI-generated text, and they accepted more AI suggestions compared to Americans. From a purely causal lens, this difference in AI engagement could be seen as a confounding factor, potentially undermining our findings. For example, it could suggest that the homogenization effects are simply a result of Indians' higher AI usage. However, we argue that this is not a flaw, but an important finding of our study. To fully understand the differential AI engagement seen in our study, we turn to HCI scholarship for a socio-technical view of human-AI interaction.

The way people perceive the benefits and harms of emerging technologies varies across cultures and geographic regions~\cite{Liu2024}. For instance, people in individualistic cultures (e.g., the US) tend to evaluate new technologies through direct and formal sources, while those in collectivistic cultures (e.g., India) rely more on peer feedback~\cite{Lee2013}. AI technologies are no different---they are cultural artifacts~\cite{Dourish2016} and perceptions about them vary across cultures, both at the national~\cite{eitle2020cultural} and user levels~\cite{Liu2024}. Research shows that users in non-Western settings express more positive emotions toward conversational agents, whereas users in the US tend to be more ambivalent~\cite{Liu2024}. Non-Western users also show higher levels of trust in AI technologies than their American counterparts~\cite{Shin2022}. In writing contexts, particularly when composing in English, AI-based writing suggestions provide more utility to non-native speakers than to native speakers, leading them to engage more with the suggestions~\cite{Buschek2021}. This positive attitude often manifests as overreliance on AI~\cite{Buinca2021}, especially among novice users in non-Western settings~\cite{Okolo2024}.

These cultural differences in AI overreliance may be exacerbated as embedded AI applications reach more diverse users. While past research has shown that overreliance can be mitigated with AI explanations~\cite{Vasconcelos2023,solano-kamaiko_explorable_2024}, this approach becomes less viable in new-age interactive and embedded AI tools, where providing explanations may not be feasible. For instance, in writing applications, presenting explanations for every suggestion may disrupt the user's flow, hindering productivity. Furthermore, our study shows that embedded AI applications promising productivity gains may inadvertently push non-Western users toward overreliance by delivering diminished benefits to them.

We argue that higher trust and AI reliance are cultural characteristics that significantly influence human-AI interaction. It is important to acknowledge these cross-cultural differences in trust, adoption, and degree of AI reliance. By trying to eliminate them in our analyses as mere confounders, we may fail to notice the cultural harms they are causing, risking a future where cultural nuances are erased.

\subsection{Harms of Cultural Homogenization}
Neocolonialism refers to the practice of using economic, political, cultural, or technological forces by Western nations to indirectly control formerly colonized nations. Unlike traditional colonialism, which involved direct territorial control, neocolonialism operates through more subtle mechanisms, such as control over resources, markets, institutions, or the spread of dominant cultural and ideological values~\cite{Scott2009, halperin2024neocolonialism}. 
In recent years, a new form of neo-colonialism has emerged, mediated by advances in big data and AI, where the digital realm is used to move wealth from the poor to the rich, a dynamic reminiscent of colonialism.~\cite{couldry2019costs}.
A growing body of scholarly work points to the harms of this \emph{data colonization}~\cite{couldry2019datacolonialism} and \emph{AI colonialism}~\cite{tacheva_ai_2023, Hao2022aicolonialism} and argues how everyday products and services offered by big tech companies based in the West capitalize on and profit from the data extracted from people in non-Western settings to enrich the wealthy and powerful. An example of
this form of colonialism is how billion-dollar tech companies such as Amazon, Facebook, Google, Microsoft, and OpenAI paying pennies to workers in developing nations to label training data~\cite{wired2024millions}, only to build products that center Western norms~\cite{Shahid2023}. In this neocolonialism, these companies establish Western influence not by political or military control, but by consolidating Western power over data, cultural expression, and information flows, inviting calls from critical scholars to decolonize computing~\cite{Irani2010, Ali2016, Mohamed2020}. 

Our study provides concrete evidence of AI colonialism that is more worrying than an AI model's internal biases: such models alter cultural preferences and descriptions of cultural artifacts \textit{even under human oversight}. This homogenization reflects a form of cultural imperialism~\cite{landsman_cultural_2014}, where one culture dominates and suppresses the plurality of knowledge, practices, and languages, reinforcing Western hegemony over values. For example, the growing use of AI in story-writing interfaces~\cite{Singh2023elephant, Lee2022, Yang2022AIAA, Yuan2022wordcraft} can gradually erode traditional storytelling methods and creative expressions. Similarly, Western-biased AI suggestions in email clients may encourage a direct and informal tone, which could lead to conflicts with superiors in cultures that observe an established social hierarchy~\cite{Hofstede2011} and expect formal language as a sign of respect. More broadly, cultures differ in directness, formality, and politeness~\cite{hershcovich-etal-2022-challenges}, distinct stylistic variations that are at risk of being lost to this homogenization. Our work highlights one such change (lexical diversity) and lays the groundwork for further research to identify more of these subtle shifts. This will require interdisciplinary effort among linguists, critical scholars, and AI researchers.

In addition to explicit cultural imperialism, the homogenization effect can also lead to an implicit \textit{cognitive} imperialism~\cite{Battiste2000-nm}. Writing is closely tied to thinking, and what one writes can transform what one thinks~\cite{Menary2007}. Continuously subjecting users to mono-cultural suggestions portrays Western norms and values as the only truth, reinforcing the supremacy of ``only one language, one culture, one frame of reference''~\cite{Battiste2000-nm}. This cultural hegemony may encourage users to think, write, and reason in ways that reflect Western values. Over time, this can affect how individuals perceive their own culture, potentially leading to feelings of cultural inferiority or a loss of cultural identity.

\subsection{Call for Mitigation Strategies}
The cultural homogenization identified in our study nudges users toward aligning their writing with Western styles. As more users accept these biased suggestions, it creates a biased feedback loop where more online content conforms to Western norms, which then becomes training data for future models, further perpetuating these biases. It is therefore critical to mitigate cultural harms at this foundational stage, as this vicious cycle could remove the opportunity to correct these issues in the future. Furthermore, current mainstream LLMs are owned by a few large companies based in the West. However, a growing trend of distilling smaller models from these larger ones to enable privacy-preserving, on-device computation~\cite{abdin2024phi3, gemmateam2024gemma2, openai2024gpt4omini} presents increased challenges in correcting these biases. For example, if cultural biases are not addressed now, they may propagate into local models and systems that are resistant to change. As a result, even if future AI systems correct these biases, legacy systems may continue to perpetuate them.

Our work makes a critical argument to \emph{design for cultural plurality}. While addressing cultural plurality is a complex challenge, as acknowledged by early studies~\cite{tamkin2021understanding, hershcovich-etal-2022-challenges}, we propose a divide-and-conquer approach. While AI researchers develop mitigation strategies at the model and architecture levels, we encourage HCI researchers and practitioners to develop solutions at the application layer. Although standard UX principles suggest that AI features be seamlessly integrated into applications so that they work out of the box, to avoid cultural harm, we need to design with \emph{friction}~\cite{mejtoft_design_2019}. This approach provides users greater control over AI features and encourages them to make appropriate configurations. For example, in writing applications, this could involve prompting the user to upload writing samples so that the AI can tailor suggestions to their personal style rather than generalizing to Western styles. More broadly, developers could provide the model with the socio-cultural context in which the application will be used or ask users to provide a personal profile to help contextualize the model's output. We invite the HCI and AI community to incorporate guardrails and intentional friction into their AI applications to design for cultural plurality and avoid the harms of cultural imperialism. 


\subsection{Limitations and Future Work}
The terms ``Western'' and ``non-Western'' are broad cultural labels, as US culture does not fully represent all Western cultures, nor does Indian culture encompass all non-Western cultures. Even within ``India'' and ``the US,'' there exist rich and diverse subcultures. In using these categorizations, we draw upon prior work~\cite{basu2023inspecting, Qadri2023regimes} that has used geographical boundaries as a proxy for cultural differences. However, future work is required to determine if our results generalize to other countries and sub-cultures. In the wake of our findings, this is especially important to understand the scale at which cultural erasure might be happening, especially for cultures that are already marginalized. Ironically, the community's focus on broad cultural categorization (e.g., treating ``India'' or ``Africa'' as monolithic cultures) may itself contribute to cultural erasure.

The scale of our study was limited by the relatively small pool of Indian participants on Prolific. Future research may expand our results by scaling horizontally across diverse cultures and vertically with larger participant pools. Further research is also needed to study the cross-cultural impact of AI suggestions on other dimensions of productivity such as cognitive load, readability, etc.

Finally, since this was a formative study, we employed a between-subjects design to explore the layers of Hofstede's Cultural Onion. This approach required analysis at the group level. Future studies could use a within-subjects design to validate whether homogenization effects persist at the individual level.

%% file: 6_conclusion.tex
We conducted a cross-cultural controlled experiment with $118$ Indian and American participants to examine the consequences of interacting with a culturally different model. Participants completed writing tasks designed to elicit cultural practices and values. They were randomly assigned to one of two conditions: the AI condition, where they received inline AI suggestions to assist with their writing, or the No AI condition, in which they wrote without any AI support. Our analysis of writing logs and essays revealed two key findings. First, AI suggestions offered greater productivity gains to American users, leaving non-Western users to put in more effort to achieve similar benefits. Second, AI homogenized writing styles towards Western norms, subtly influencing Indians to adopt Western writing patterns.
This homogenization provides concrete evidence of AI colonialism, where these models reinforce Western cultural hegemony.
It is crucial to address these biases before culturally biased content proliferates and becomes part of the training data for future AI models.

%% file: 7_appendix.tex
\section{Participant Demographic Details} \label{appendix:participant_demographics}
Please refer Table~\ref{tab:languages_occupations}.

\begin{table*}[ht]
\centering
\begin{tabular}{|l|l|l|}
\hline
 & \textbf{Languages} & \textbf{Occupations} \\ \hline
\textbf{India} & \begin{tabular}[c]{@{}l@{}}Bengali, English, French,\\ Garhwali, German, Gujarati,\\ Hindi, Kannada, Kashmiri,\\ Konkani, Malayalam, Urdu,\\ Nepali, Marathi, Punjabi,\\ Sanskrit, Tamil, Telugu\end{tabular} & \begin{tabular}[c]{@{}l@{}}Academician/Professor, Accountant/Finance, Doctor,\\ Business Professional, Counsellor, Data Analyst,\\ Designer, Employee (General), IT Professional,\\ Developer/Software Engineer, Entrepreneur, Intern,\\ Freelancer/Consultant, Student, Self-employed,\\ Teacher/Tutor, Manager (Marketing/Sales), Writer\end{tabular} \\ \hline
\textbf{US} & \begin{tabular}[c]{@{}l@{}}Chinese, English, French,\\ German, Indonesian, Italian,\\ Japanese, Korean, Spanish,\\ Vietnamese\end{tabular} & \begin{tabular}[c]{@{}l@{}}Academician, Construction Worker, Consultant,\\ Cook, Copywriter, Data Analyst, Driver, Teacher,\\ Developer/Software Engineer, Finance, Artist,\\ Business Professional, Employee (General),\\ Health Care Worker, Homemaker, IT Professional,\\ Manager (Marketing/Sales), Military, Real Estate,\\ Project Engineer, Retail Worker, Retired, Security,\\ Self-Employed, Student, Transcriber, Transportation,\\ Unemployed, Veterinary Assistant, Video Editor\end{tabular} \\ \hline
\end{tabular}
\caption{Full list of languages and occupations self-disclosed by our participants}
\Description{A table listing the full set of languages and occupations self-disclosed by participants from India and the US. Languages: Indian participants reported speaking 18 languages, including Bengali, Hindi, Tamil, and Urdu, alongside English and several other international languages. US participants reported speaking 10 languages, including Chinese, Spanish, Korean, and English. Occupations: Both Indian and American participants reported diverse roles, including Academician/Professor, IT Professional, Data Analyst, Entrepreneur, Freelancer/Consultant, and Teacher/Tutor.}
\label{tab:languages_occupations}
\end{table*}

\section{Favorite Celebrities Selected by Participants} \label{appendix:celebrities_selected}
Please refer Table~\ref{tab:celebrities_selected}.
\input{tables/celebrities_selected}

\section{Changing Cultural Expression: Additional Examples} \label{appendix:essay_examples_appendix}
Please refer Table~\ref{tab:essay_examples_appendix}.
\input{tables/essay_examples_appendix}

%% file: tables/celebrities_selected.tex
\begin{table*}[t]
\centering
\renewcommand{\arraystretch}{1.3}
\begin{tabular}{|l|c|l|c|}
\hline
\textbf{No AI} ($n$=24)                 & \textbf{Count} & \textbf{AI} ($n$=36)                    & \textbf{Count} \\ \hline
\cellcolor[HTML]{D8EAF2} Shah Rukh Khan & 3              & \cellcolor[HTML]{D8EAF2} Shah Rukh Khan & 5              \\ 
\cellcolor[HTML]{D8EAF2} Virat Kohli    & 2              & \cellcolor[HTML]{D8EAF2} APJ Abdul Kalam & 4              \\ 
\cellcolor[HTML]{D8EAF2} Rajinikanth    & 2              & \cellcolor[HTML]{D8EAF2} Priyanka Chopra & 1              \\ 
\cellcolor[HTML]{D8EAF2} Lakshay Sen    & 1              & \cellcolor[HTML]{D8EAF2} Mahatma Gandhi  & 1              \\ 
\cellcolor[HTML]{D8EAF2} Konidela Pavan Kalyan & 1       & \cellcolor[HTML]{D8EAF2} Diljit Dosanjh  & 1              \\ 
\cellcolor[HTML]{D8EAF2} Kangana Ranaut & 1              & \cellcolor[HTML]{D8EAF2} Amitabh Bachchan & 1              \\ 
\cellcolor[HTML]{D8EAF2} Salman Khan    & 1              & \cellcolor[HTML]{D8EAF2} Allu Arjun      & 1              \\ 
\cellcolor[HTML]{D8EAF2} Arijit Singh   & 1              & \cellcolor[HTML]{D8EAF2} Alia Bhatt      & 1              \\ 
\cellcolor[HTML]{D8EAF2} P Sainath    & 1              & \cellcolor[HTML]{D8EAF2} A.R. Rahman     & 1              \\ 
\cellcolor[HTML]{FFF4E0} Warren Buffet  & 1              & \cellcolor[HTML]{D8EAF2} Yogi Adityanath & 1              \\ 
\cellcolor[HTML]{FFF4E0} Tom Hardy      & 1              & \cellcolor[HTML]{D8EAF2} Vishwanathan Anand & 1          \\ 
\cellcolor[HTML]{FFF4E0} Terrence Tao   & 1              & \cellcolor[HTML]{D8EAF2} Virat Kohli     & 1              \\ 
\cellcolor[HTML]{FFF4E0} Sebastian Vettel & 1            & \cellcolor[HTML]{D8EAF2} Vinesh Phogat   & 1              \\ 
\cellcolor[HTML]{FFF4E0} Barack Obama   & 1              & \cellcolor[HTML]{D8EAF2} Samay Raina     & 1              \\ 
\cellcolor[HTML]{FFF4E0} Miley Cyrus    & 1              & \cellcolor[HTML]{D8EAF2} Sadh Guru       & 1              \\ 
\cellcolor[HTML]{FFF4E0} Kim Namjoon    & 1              & \cellcolor[HTML]{D8EAF2} Sachin Tendulkar & 1            \\ 
\cellcolor[HTML]{FFF4E0} Elon Musk      & 1              & \cellcolor[HTML]{FFF4E0} Sylvester Stallone & 2          \\ 
\cellcolor[HTML]{FFF4E0} Cristiano Ronaldo & 1           & \cellcolor[HTML]{FFF4E0} Ryan Reynolds   & 2              \\ 
\cellcolor[HTML]{FFF4E0} Zayn Saifi     & 1              & \cellcolor[HTML]{FFF4E0} Cristiano Ronaldo & 2          \\ 
\cellcolor[HTML]{FFF4E0} Emma Watson    & 1              & \cellcolor[HTML]{FFF4E0} Selena Gomez    & 1              \\ 
                                       &                & \cellcolor[HTML]{FFF4E0} Keanu Reeves    & 1              \\ 
                                       &                & \cellcolor[HTML]{FFF4E0} Rebecca Solnit  & 1              \\ 
                                       &                & \cellcolor[HTML]{FFF4E0} Leon Marchand   & 1              \\ 
                                       &                & \cellcolor[HTML]{FFF4E0} Christo Xavier   & 1              \\ 
                                       &                & \cellcolor[HTML]{FFF4E0} Chamath Palihapitiya & 1          \\ 
                                       &                & \cellcolor[HTML]{FFF4E0} Adam Khoo       & 1              \\ \hline
\end{tabular}
\caption{Public figures chosen by Indian participants across the AI and No AI conditions. 
\colorbox[HTML]{D8EAF2}{Blue cells} indicate Indian public figures, 
\colorbox[HTML]{FFF4E0}{yellow cells} represent non-Indian public figures.}
\label{tab:celebrities_selected}
\Description{Blue cells indicate Indian public figures. In both conditions, Shah Rukh Khan was the most frequently chosen Indian figure. Yellow cells represent non-Indian public figures chosen by Indian participants. Warren Buffet, Tom Hardy, and Elon Musk, among others, were chosen in the No AI condition, and Sylvester Stallone, Ryan Reynolds appeared in the AI condition.}
\end{table*}

%% file: tables/essay_examples_appendix.tex
\begin{table*}[t]
\small
\centering
\begin{tabular}{l|ll}
\multicolumn{1}{c|}{\textbf{Topic}} & \multicolumn{1}{c}{\textbf{Essay without AI}} & \multicolumn{1}{c}{\textbf{Essay with AI}} \\ \hline
\begin{tabular}[c]{@{}l@{}}Festival\\ (Onam)\end{tabular} & \begin{tabular}[c]{@{}l@{}}My favorite festival is Onam which is a festival\\ celebrated in my home state, Kerala. One of the\\ specialties of Onam is the sadya. In northern\\ Kerala Onam sadya includes non- vegetarian\\ dishes and in \colorbox{PastelGreen}{central and southern Kerala} it is\\ mainly a vegetarian affair. I usually prepare the\\ non-vegetarian sadya along with \colorbox{PastelGreen}{two types of}\\ \colorbox{PastelGreen}{payasam}. We also wear new clothes on the day\\ called \colorbox{PastelGreen}{"onakkodi"}. Onam is a harvest festival.\end{tabular} & \begin{tabular}[c]{@{}l@{}}I celebrate onam which is a state festival of kerala.\\ \textbf{We decorate our homes with flowers}. and prepare\\ traditional meals \textbf{called Onasadya.} \textbf{We also wear}\\ \textbf{traditional attire.} The amain highlight is get-\\ together with family and friends \colorbox{PastelYellow}{\textbf{and participating}}\\ \colorbox{PastelYellow}{\textbf{in cultural events}.} In every nook and corner of the\\ state there will cultural events and games which can\\ be enjoyed without any discrimination\\ and \textbf{with }\colorbox{PastelYellow}{\textbf{great enthusiasm.}}\end{tabular} \\ \hline
\begin{tabular}[c]{@{}l@{}}Festival\\ (Pongal)\end{tabular} & \begin{tabular}[c]{@{}l@{}}My Favorite Festival is Pongal which is\\ celebrated in January. We celebrate with our\\ family relatives for 3 days- Boggi, Kanuma,\\ Makara Sankranti. 1st Boggi is celebrated\\ \colorbox{PastelGreen}{abound firewood} with relatives and neighbours\\ and then kanuma Sankranthi which is celebrated\\ by \colorbox{PastelGreen}{drawing colours on the grounds} and visiting\\ the temple and Makara Sankranthi which we\\ eat together and enjoy together.\end{tabular} & \begin{tabular}[c]{@{}l@{}}My cultural festival is Pongal.  We celebrate this\\ festival to \textbf{honor the sun god and the harvest.}\\ We have three days to celebrate three pongal \textbf{days:}\\ \textbf{Bhogi, Surya, and Mattu Pongal.} We celebrate\\ with \colorbox{PastelYellow}{\textbf{delicious food, family gatherings, and}}\\ \colorbox{PastelYellow}{\textbf{traditional dances}.} All wear new \textbf{clothes and}\\ \colorbox{PastelYellow}{\textbf{decorate their homes}.} We clean \textbf{our homes}\\ \textbf{thoroughly and draw beautiful kolams.}\end{tabular} \\ \hline
\begin{tabular}[c]{@{}l@{}}Festival\\ (Ugadi)\end{tabular} & \begin{tabular}[c]{@{}l@{}}My favorite festival is Ugadi. Being the first year\\ for Telugu people, it is a major festival. We used\\ to wear new dress and prepare and \colorbox{PastelGreen}{consume}\\ \colorbox{PastelGreen}{Ugadi Pachhadi}. This helps me to identify and\\ accept both \colorbox{PastelGreen}{difficulties and happiness are part of}\\ \colorbox{PastelGreen}{life} and need to accept them as a part. Also there\\ \colorbox{PastelGreen}{is a panchanga shravanam} where I listen to this to\\ find how my future looks in the present year.\end{tabular} & \begin{tabular}[c]{@{}l@{}}My favorite festival is Ugadi because it is new year\\ day \textbf{for many South Indian states.} I ce\textbf{lebrate it with}\\ \textbf{traditional dishes and family gatherings.} I \textbf{also enjoy}\\ \colorbox{PastelYellow}{\textbf{decorating the house with flowers}.} I get up early and\\ take an head bath wear new clothes and do puja or\\ anustana at my home. \colorbox{PastelYellow}{\textbf{with devotion and joy}}. I offer \\ \textbf{prayers and }\colorbox{PastelYellow}{\textbf{special dishes to the deities}}. A special\\ portion called Ugadi Pachadi is prepared and offered\\ first to my presiding deities \textbf{to seek their blessings for}\\ \textbf{the new year.} Then it will be consumed \textbf{by everyone}\\ \textbf{in the family as a ritual.} It \textbf{marks the beginning of a}\\ \textbf{prosperous year for us.}\end{tabular} \\ \hline
\begin{tabular}[c]{@{}l@{}}Food\\ (Dosa)\end{tabular} & \begin{tabular}[c]{@{}l@{}}My favorite food is \colorbox{PastelGreen}{ghee masala dosa}, a South\\ Indian cuisine. I like it because it is very tasty and\\ healthy. Firstly, it is made with clarified butter\\ which is healthy for digestion. Secondly, it is\\ accompanied with \colorbox{PastelGreen}{lentil soup called sambar} which\\ has lots of yummy flavor and a good mix of healthy\\ veggies. It is also accompanied with a number of\\ chutneys, which are a \colorbox{PastelGreen}{paste of gram dal , coconut,}\\ \colorbox{PastelGreen}{tomato , onion , garlic and coriander} . It is very\\ delicious and healthy. Finally. I like ghee masala\\ dosa because it is easy to digest and at the dame\\ time full of gastronomic flavors.\end{tabular} & \begin{tabular}[c]{@{}l@{}}My favorite food is Dosa  \textbf{because it's crispy and}\\ \textbf{delicious.} It can be filled with different ingredients \textbf{like}\\ \textbf{potatoes or }\colorbox{PastelBlue}{\textbf{cheese.}} It's nutritious, easy to digest, low in\\ fat and culturally significant, making it a popular choice\\ for many \colorbox{PastelBlue}{\textbf{people in India and beyond}.} D\textbf{osa is also}\\ \textbf{versatile and can be enjoyed with various chutneys}\\ \textbf{and sambar.}\end{tabular}
\end{tabular}
\caption{Additional examples from the content analysis comparing cultural artifacts described by Indian participants with and without AI suggestions. AI-generated text is shown in \textbf{bold}. Highlights indicate key differences: \colorbox{PastelGreen}{green} for cultural details present without AI but missing with AI, \colorbox{PastelYellow}{yellow} for generic phrasing introduced by AI, \colorbox{PastelBlue}{blue} for Westernized descriptions.}
\Description{Four topics are shown: "Festival (Onam)," "Festival (Pongal)," "Festival (Ugadi)," and "Food (Dosa)." For festivals, the essays without AI include specific cultural details such as regional variations in Onam meals, firewood traditions in Pongal, and the symbolic meaning of Ugadi Pachadi, while the AI-assisted essays emphasize more generic descriptions like "traditional dishes," "family gatherings," and "decorating the house." For "Food (Dosa)," the essay without AI describes preparation details, including ghee, lentil soup (sambar), and specific chutney ingredients, while the AI-assisted version provides Westernized descriptions.}

\label{tab:essay_examples_appendix}
\end{table*}